# Evolutionary characterization of whole-genome doubling across human cancers

Yangyang Zhang[1,2,†], Haiming Li[2,3,†], Xiaoyu Zhao[4,†], Zilong Wang[5,6], Yongqi Chen[1,2], Sumeng Yang[1,2], Shaocong Mo[7], Jinming Li[2,8], Xiaohua Wu[1,2,*], Xu Wang[2,9,10,11,12,13,*], Shuo Wang[14,15,*], Qinhao Guo[1,2,*]

[1]Department of Gynecologic Oncology, Fudan University Shanghai Cancer Center, Fudan University, Shanghai 200032, China.

[2]Department of Oncology, Shanghai Medical College, Fudan University, Shanghai 200032, China.

[3]Department of Radiology, Fudan University Shanghai Cancer Center, Fudan University, Shanghai 200032, China.

[4]Department of Medicine and Moores Cancer Center, University of California, San Diego, La Jolla, CA 92093, USA.

[5]Microsoft Research Asia, Shanghai 200232, China.

[6]CataX AI, 900 Yishan Road, Shanghai 200233, China.

[7]Department of Digestive Diseases, Huashan Hospital, Fudan University, 12 Wulumuqi Middle Road, Shanghai 200040, China.

[8]Department of Colorectal Surgery, Fudan University Shanghai Cancer Center, Shanghai 200032, China.

[9]Cancer Research Institute, Fudan University Shanghai Cancer Center, Fudan University, Shanghai 200032, China.

[10]Department of Pancreatic Surgery, Fudan University Shanghai Cancer Center, Shanghai 200032, China.

[11]Shanghai Pancreatic Cancer Institute, Shanghai 200032, China.

[12]Shanghai Key Laboratory of Precision Medicine for Pancreatic Cancer, Shanghai 200032, China.

[13]Pancreatic Cancer Institute, Fudan University, Shanghai 200032, China.

[14]Digital Medical Research Centre, School of Basic Medical Sciences, Fudan University, Shanghai 200032, China.

[15]Shanghai Key Laboratory of MICCAI, Shanghai 200032, China.

[†]These authors contributed equally.

*Corresponding authors.

E-mail addresses: wu.xh@fudan.edu.cn (Xiaohua Wu), wangxu2013@fudan.edu.cn (Xu Wang), shuowang@fudan.edu.cn (Shuo Wang), guoqinhao_sh@fudan.edu.cn (Qinhao Guo).

**Abstract**

Whole-genome doubling (WGD) is common across human cancers and is associated with tumour progression and adverse clinical outcomes.[1-4] However, how genome-doubled tumours evolve across diverse cancer types remains poorly understood. Here we characterize WGD-associated tumour evolution by integrating genomic analyses of 41,332 tumours from 12 cancer types with evolutionary timing and multi-omic data. Several canonical driver mutations generally preceded WGD when both occurred, as did the onset of homologous recombination deficiency-associated mutagenesis. Notably, post-WGD chromosome-arm losses showed substantially weaker cancer-type dependence than pre-WGD losses, with 61–75% less variance explained by cancer type. Beyond chromosomal evolution, WGD+ tumours showed reduced interferon-response and HLA antigen-presentation programmes linked to epigenetic remodelling, without disproportionate enrichment of genetic immune-escape alterations. Our findings highlight WGD as a recurrent evolutionary transition associated with altered chromosomal evolution and epigenetically linked immune adaptation across cancers.

## Introduction

Whole-genome doubling (WGD), an event that doubles a tumour cell's entire chromosome complement, occurs in approximately one-third of human solid tumours. It is associated with chromosomal instability, metastatic disease and poor clinical outcome.[1,2] By increasing genomic redundancy, doubling may buffer deleterious chromosome losses and broaden the range of karyotypes that can persist.[3,4] Previous studies have shown that several driver mutations often precede WGD, whereas chromosomal alterations continue to accumulate afterwards.[1,3-6] However, the evolutionary timing of WGD-associated genomic alterations and the cancer-type specificity of subsequent chromosomal evolution remain incompletely characterized.

Single-cell studies have revealed ongoing and parallel genome doubling within tumours,[7] and metastatic phylogenies have documented additional doubling after dissemination.[8] Beyond these genomic changes, experimental studies have linked WGD to chromatin reorganization[9] and, in breast cancer models, to epigenetic suppression of antigen presentation.[10] It remains unclear whether similar immune alterations extend across cancer types and how they relate to epigenetic remodelling and post-WGD genetic immune-escape alterations.

Here we systematically characterize WGD as an evolutionary transition across human cancers using genomic data from 41,332 tumours spanning 12 cancer types, together with evolutionary timing, clinical and multi-omic analyses. We find that post-WGD chromosome-arm losses show substantially weaker cancer-type dependence than pre-WGD losses, despite diverse preceding genetic histories. WGD+ tumours also exhibit reduced interferon-response and HLA antigen-presentation programmes linked to epigenetic remodelling, without disproportionate enrichment of genetic immune-escape alterations.

## Results

### Prevalence and correlates of WGD

We assembled 41,332 tumours from 37,833 patients across 12 cancer types, including 33,968 tumours profiled by targeted sequencing in MSK-IMPACT and 7,364 primary tumours profiled by genome-wide SNP arrays in TCGA (**Fig. 1a,b** and **Supplementary Table 1**). WGD was defined as a major copy number of at least two across more than 50% of the autosomal genome. The genomic fraction meeting this criterion showed a bimodal distribution in both cohorts (**Fig. 1c** and **Extended Data Fig. 1a**). Overall, 14,304 tumours (34.6%) were WGD+, with prevalence ranging from 19.8% in prostate cancer to 53.9% in ovarian cancer (**Fig. 1d**). WGD frequencies were strongly correlated between cohorts across cancer types (Spearman $\rho = 0.87$), despite being consistently higher in TCGA than in MSK-IMPACT (46.0% versus 32.1% overall; **Fig. 1e**).

WGD was strongly associated with *TP53* mutation across all 12 cancer types. Among 40,730 tumours with mutation data, *TP53* mutations were present in 69.7% of WGD+ tumours compared with 43.0% of WGD− tumours (cancer-type-stratified odds ratio (OR), 2.95; 95% confidence interval (CI), 2.82–3.10; $q < 0.05$ in each cancer type; **Fig. 1f**). Homologous recombination deficiency (HRD) scores were also higher in WGD+ tumours in 11 of 12 cancer types (6,056 TCGA tumours; paired Wilcoxon $P = 0.021$; **Fig. 1g**). Ovarian cancer was the exception, with median HRD scores of 41 in WGD+ tumours and 57 in WGD− tumours (**Supplementary Table 1**).[11]

WGD was associated with shorter overall survival among 26,184 MSK-IMPACT patients, with median survival from first sequencing of 31 months for WGD+ tumours versus 53 months for WGD− tumours (hazard ratio (HR), 1.30; 95% CI, 1.24–1.35; stratified by cancer type and disease state at sequencing; **Fig. 1h**). The association was directionally consistent across all 12 cancer types and persisted after adjustment for *TP53* mutation (HR, 1.20; 95% CI, 1.15–1.25; **Extended Data Fig. 1b,c**). An independent analysis of 1,440 HRD-tested patients with stage III–IV high-grade serous ovarian cancer in FUSCC-ClinSeq confirmed the association with shorter survival (HR, 1.67; 95% CI, 1.31–2.14), which persisted after adjustment for age, FIGO stage, HRD status, CA-125, residual disease and neoadjuvant chemotherapy (HR, 1.51; 95% CI, 1.18–1.95; P = 0.0012; **Fig. 1i** and **Supplementary Table 2**).

**Genomic associations with WGD**

We mapped WGD-associated mutations, focal copy-number alterations and chromosome-arm LOH in primary tumours from MSK-IMPACT (n = 12,190) and TCGA (n = 7,355), using a null model accounting for cancer type, microsatellite status and alteration burden (**Fig. 2a** and **Methods**). Simulation-based calibration and silent-mutation controls supported these tests (**Extended Data Fig. 2a–d**).

We identified 67 distinct WGD-associated events with concordant directions across cohorts: 39 in the pan-cancer analysis and 57 in at least one cancer type, with 29 shared between analyses (**Fig. 2b**). These represented 113 retained associations when cancer-type-specific findings were counted separately. Mutual exclusivity was more frequent than co-occurrence (39 versus 28 events; **Fig. 2c**), and cancer-type-specific associations generally showed larger effect sizes (**Fig. 2d**). Of 70 associations evaluable in independent ICGC/PCAWG genomes, 20 met a false discovery rate (FDR) threshold of 0.10, all concordant with the discovery direction (**Extended Data Fig. 2e,f** and **Supplementary Table 3**).

*TP53* was the only mutated gene showing pan-cancer co-occurrence with WGD (enrichment ratio (ER; observed/expected), 1.18 in MSK-IMPACT and 1.15 in TCGA), with the association retained in seven cancer types after null-model adjustment (**Fig. 2e**). By contrast, mutations in ten driver genes, including *ARID1A*, *KRAS*, *PIK3CA* and *PTEN*, were mutually exclusive with WGD. Copy-number associations included co-occurrence of *CCNE1* amplification (ER, 1.36 and 1.20) and 17p LOH, and mutual exclusivity of 1q gain and 16q LOH (**Fig. 2e** and **Supplementary Table 3**).

The depleted drivers differed across cancer types, including *CDH1*, *GATA3* and *PIK3CA* in breast cancer; *KRAS* and *GNAS* in colorectal cancer; *FGFR3* and *HRAS* in bladder cancer; and *PTEN* and *CTCF* in endometrial cancer (**Fig. 2e**). Cancer-type-specific co-occurrence included E2F3 amplification in bladder cancer and PPP2R1A mutation in endometrial cancer.

To determine whether lineage-specific driver associations involved shared protein assemblies, we analysed 87 testable assemblies from the NeST hierarchy (**Extended Data Fig. 3a**).[12,13] Fourteen showed concordant associations across cohorts: six were depleted in WGD+ tumours, including RAS–RAF–MAPK and TGF-β–SMAD assemblies, whereas eight were enriched, principally involving DNA repair and TP53 regulation (**Extended Data Fig. 3b**). These analyses identified shared and lineage-specific genomic associations with WGD but did not establish their temporal ordering.

## Timing of genomic events around WGD

To place WGD-associated genomic alterations in evolutionary context, we reconstructed event timing relative to doubling using mutation multiplicity and allele-specific copy number in MSK-IMPACT (10,071 WGD+ patients), TCGA (2,183 WGD+ tumours) and a quality-filtered PCAWG subset (417 WGD+ tumours; **Fig. 3a** and **Methods**).

HRD-associated mutagenesis generally preceded WGD. Among 95 PCAWG tumours with estimable onset of the HRD-associated signature SBS3, 86 showed onset before doubling, with the inferred onset occurring a median of 0.37 earlier on a relative mutation-time scale (0–1) (**Fig. 3b**). The same ordering was observed in all 31 HRD-positive WGD+ high-grade serous ovarian cancers analysed in FUSCC-OV (**Extended Data Fig. 4a**).

Driver mutations also predominantly preceded WGD. In MSK-IMPACT, 15 of 17 frequently mutated driver genes had pre-WGD fractions exceeding the 70% baseline among clonal non-synonymous mutations (**Fig. 3c**). *TP53* mutations preceded WGD in 94% of timed cases in both MSK-IMPACT and TCGA, while *STK11*, *KEAP1*, *CDKN2A* and *KRAS* preceded doubling in 87–92% of cases in MSK-IMPACT. Across cancer types and platforms, 16 of 17 drivers showed predominantly pre-WGD timing, with *PIK3R1* unresolved (**Fig. 3d** and **Supplementary Table 4**). Significant timing heterogeneity across cancer types was limited to *TP53* and *PIK3CA*, and predominantly pre-WGD driver timing was reproduced in TRACERx lung-cancer exomes (**Extended Data Fig. 4b**). Notably, *KRAS*, *PIK3CA* and *PTEN* mutations generally preceded doubling when present, despite being depleted in WGD+ tumours.

Chromosome-arm losses, in contrast, occurred both before and after WGD, with timing differing across cancer types for all nine recurrent losses examined. Using allele-specific copy-number states independently of the enrichment null, we found that 17p loss predominantly preceded WGD (probability, 0.78; 95% CI, 0.73–0.83), whereas losses of 6q, 17q, 11p, 12q, 20p and 3q predominantly followed it; 16q and 22q remained unresolved (**Fig. 3d**). Across 39 chromosome arms, pre-WGD fractions were strongly correlated with estimates from Hartwig metastatic genomes and TRACERx multi-region exomes (Spearman $\rho = 0.93$ and 0.87, respectively; **Extended Data Fig. 4c,d**).

Single-cell phylogenies further supported these timing estimates and revealed lineage-specific differences. In 25 WGD-containing clone trees from high-grade serous ovarian cancer, 17p LOH preceded doubling in every tree (**Fig. 3e**). Timing estimates for nine chromosome-arm losses closely matched ovarian bulk estimates (Spearman $\rho = 0.98$), including pre-WGD losses of 17q and 22q despite their later or unresolved pan-cancer timing (**Extended Data Fig. 4f**). Additional clone phylogenies across ovarian, breast, colorectal, renal and lung cancers identified pre-existing arm-level alterations before WGD (**Extended Data Fig. 5a–e**).[14-16]

Chromosome gains also continued after WGD. Published gain-timing posteriors from 492 WGD+ PCAWG tumours showed that, across 91 chromosome-arm–cancer-type combinations, the mean probability that the first gain preceded WGD was 0.53, whereas the mean probability that the last gain followed WGD was 0.81 (**Fig. 3f**). The latter exceeded 0.5 in all 91 groups, including arms with WGD-associated gains. Similar patterns were observed in Hartwig genomes (**Extended Data Fig. 4e**).

We next asked whether genomic alterations remained as distinct across cancer types after WGD as they were before it. Using a shared event set and matched pre- and post-WGD event counts in 7,313 MSK-IMPACT tumours, all ten cancer types showed reduced lineage distinctiveness after doubling (**Fig. 3g**). Cancer type explained 1.32% of variation in event patterns before WGD but only 0.54% afterwards (59% reduction; permutation $P < 0.0005$), with a similar reduction in TCGA (3.14% to 0.86%; 73%).

This reduction was particularly consistent for chromosome-arm losses, reaching 61%, 75% and 66% in MSK-IMPACT, TCGA and Hartwig metastatic genomes, respectively. Differences between cancer-type averages narrowed, whereas variation within cancer types did not decrease (**Supplementary Table 5**). Consistent with this, cancer-type prediction accuracy declined after WGD, from 22.5% to 15.6% in MSK-IMPACT and from 43.0% to 24.4% in TCGA. At the individual-event level, lineage dependence decreased for 31 of 39 chromosome-arm losses (median change, −65%; $P = 1.6 \times 10^{-5}$), but not significantly for driver mutations (31 of 55; median change, −15%; $P = 0.082$; **Supplementary Table 6**). Together, these findings reveal weaker cancer-type dependence of chromosome-arm losses after WGD, despite the distinct genetic histories preceding doubling.

**WGD and metastatic progression**

The distinct timing of driver mutations and chromosome-arm losses raised the question of whether this evolutionary ordering persists during metastasis and how WGD relates to metastatic risk. In MSK-IMPACT, WGD prevalence increased from 24.1% in primary tumours from patients with localized disease to 31.9% in primary tumours from patients with metastatic disease and 39.2% in metastases, with a significant trend in ten of 12 cancer types (**Fig. 4a**). WGD was likewise more frequent in Hartwig metastases than in PCAWG primary tumours (56.6% versus 34.8%), with enrichment across sampled metastatic sites (**Extended Data Fig. 6a**).

WGD in primary tumours was also associated with subsequent metastasis. Among 4,925 MSK-MET patients without metastasis at primary-tumour surgery, three-year metastasis incidence was 51.7% for WGD+ tumours versus 40.5% for WGD− tumours (cancer-type-stratified hazard ratio (HR), 1.41; 95% confidence interval (CI), 1.27–1.56; **Fig. 4b**). In FUSCC-ClinSeq, miliary dissemination, a diffuse pattern of peritoneal spread, was more frequent in WGD+ than WGD− high-grade serous ovarian cancers among 1,440 patients with stage III–IV disease (57.5% versus 19.5%; **Fig. 4c**).

We next compared WGD-associated genomic alterations between primary and metastatic tumours. Across repeated MSK-IMPACT comparisons matched by cancer type and sample size, 152 of 395 loci (38.5%) showed differences in association strength exceeding the 99th percentile of matched-proportion controls (**Fig. 4d,e**). Associations with 8p and 14q loss of heterozygosity (LOH) were more strongly positive in metastases, whereas 9p LOH showed stronger depletion in WGD+ primary tumours. Despite these differences, WGD-associated effect sizes remained strongly correlated between primary and metastatic tumours (Spearman $\rho = 0.79$), with no reversals among 79 recurrently associated loci (**Fig. 4f**). Among eleven alterations associated with metastatic status after adjustment for WGD and other covariates, 8p LOH was also positively associated with metastasis (OR, 1.42; 95% CI, 1.27–1.58). However, seven of these eleven alterations showed negative WGD-association estimates, while WGD itself remained associated with metastatic status (OR, 1.93; 95% CI, 1.70–2.19; **Fig. 4g**). Thus, WGD and metastatic disease shared some genomic correlates, but their associated alterations were not identical.

We next examined WGD detection in 1,036 matched primary–metastasis pairs and used published multiregion phylogenies to reconstruct its timing relative to metastatic seeding. WGD was detected at both sites in 294 pairs, only in the metastasis in 211 and only in the primary tumour in 121 (**Fig. 4h,i**). Metastasis-only detection remained more frequent after adjustment for tumour purity (ratio, 1.61; 95% CI, 1.28–2.03). This discordance could reflect WGD arising after dissemination or the expansion of pre-existing doubled subclones missed during primary-tumour sampling. In multi-region autopsy phylogenies from 22 patients with non-small-cell lung cancer, WGD was inferred to precede metastatic seeding in 16 patients, follow seeding in four and occur only in primary-tumour lineages in two (**Fig. 4j** and **Extended Data Fig. 6b**). These phylogenies also revealed pre-WGD driver and chromosomal alterations alongside additional alterations acquired on metastatic branches (**Extended Data Fig. 6c**).

Finally, we examined whether metastatic evolution preserved the ordering of genomic events around WGD. Pre-WGD fractions for driver mutations and chromosome-arm losses were strongly correlated between unpaired WGD+ primary and metastatic tumours (Spearman $\rho$ = 0.91; **Extended Data Fig. 6d** and **Supplementary Table 4**). Among the 294 pairs with WGD at both sites, metastases tended to harbour more private post-WGD events than their paired primary tumours (mean, 2.28 versus 1.96; paired Wilcoxon $P = 0.054$). In 289 pairs with events timed in both samples, 2,457 of 2,751 jointly timed events (89.3%) retained their placement before or after WGD (Cohen's $\kappa = 0.79$; **Extended Data Fig. 6e**). Thus, metastatic genomes continued to accumulate alterations while largely retaining the evolutionary ordering established around WGD.

## WGD and tumour immunity

We next asked whether genome-doubled tumours also share molecular programmes across cancer types despite their distinct genetic histories. Across 8,951 TCGA transcriptomes, 976 CPTAC proteomes and 865 phosphoproteomes, immune and interferon-response programmes were among the most reduced in WGD+ tumours, whereas proliferation programmes were increased (**Fig. 5a** and **Supplementary Table 7**). Immune-programme scores were lower in 11 of 12 cancer types (**Supplementary Table 8**). An independent ovarian cancer cohort showed reduced IFN-γ response at both RNA and protein levels and reduced IFN-α response at the RNA level (**Extended Data Fig. 7a,b**). FUSCC-OV metabolomes also identified two pathways, including purine metabolism, with lower metabolite abundances (**Extended Data Fig. 7c**).

As bulk tumour profiles can reflect differences in cell composition, we normalized immune features to the immune compartment and accounted for proliferation within each cancer type. Among 5,847 TCGA tumours, 68 of 162 features differed by WGD status (empirical FDR < 0.05; **Fig. 5b** and **Supplementary Table 9**), with lower T-cell-associated and higher macrophage-associated signals. We then analysed single-cell transcriptomes from 76 patients with high-grade serous ovarian cancer in SPECTRUM and FUSCC-OV. Cancer cells from 36 WGD-high or WGD+ tumours showed lower MHC class I antigen-presentation, IFN-γ-response and IFN-α-response scores than those from 40 WGD-low or WGD− tumours (q = 0.010 for each; **Fig. 5c**). Expression was lower for 9 of 14 MHC class I genes, 5 of 8 class II genes and *STING1*, whereas the MHC class II programme and interferon-signalling genes showed no significant overall differences (**Fig. 5d**). Chromatin-accessibility-derived gene activity was also lower for 12 of 14 MHC class I genes in breast and colorectal cancer cells (**Extended Data Fig. 7f**).

We next examined clinical outcomes among 1,004 patients across seven cancer types treated with immune checkpoint blockade. Patients with WGD+ tumours (n = 366) had shorter overall survival after adjustment for tumour mutational burden and clinical covariates (HR, 1.25; 95% CI, 1.04–1.50; P = 0.018; **Fig. 5e**), with similar estimates across alternative models (HR, 1.19–1.25; **Fig. 5f**).

Across cell lines, organoids and isogenic mouse tumours, WGD-associated immune changes varied by experimental context: differences were not significant in cell lines or organoids, and the nominal reductions seen in immunocompetent mice at day 14 did not withstand multiple-testing correction (**Fig. 5g**). RNA-level differences were broader than those inferred from chromatin accessibility (**Extended Data Fig. 7g**).

We also assessed genetic alterations affecting immune recognition. HLA class I loss of heterozygosity (LOH) was more frequent in WGD+ tumours (19.6% versus 12.5%; n = 6,880), but showed no additional enrichment relative to genome-wide LOH burden (P = 0.79; **Fig. 5h**). Truncating mutations across six immune-escape gene sets showed no consistent excess among 6,266 non-hypermutated tumours (**Supplementary Table 10**).

We next examined DNA methylation as a potential correlate of the reduced immune programmes. Across 5,139 tumours from 12 cancer types, WGD+ tumours showed lower methylation within partially methylated domains in ten cancer types (P = 0.002; **Fig. 5i** and **Supplementary Table 11**), alongside local hypermethylation enriched at PRC2-target regions in 11 cancer types (P = 0.00098; **Fig. 5j**). PRC2 targets accounted for 54% of CpG islands hypermethylated with WGD in at least two cancer types, compared with 33% of measured islands (**Extended Data Fig. 7d** and **Supplementary Table 12**). In isogenic tumour data, genes with PRC2-target promoter islands were also enriched in regions shifting towards inactive chromatin (OR, 1.86; $P = 1.4 \times 10^{-8}$; **Extended Data Fig. 7e**).

Notably, allograft rejection and IFN-γ response, the programmes reduced at the transcriptomic level, were also the most enriched among WGD-associated hypermethylated regions (**Fig. 5k**). Promoter hypermethylation was more frequent in WGD+ tumours for 29 of 32 immune genes, reaching significance for 13 (**Fig. 5l**). Across 4,116 tumours with matched methylation and expression data, larger WGD-associated increases in promoter hypermethylation correlated with greater expression reductions associated with hypermethylation, after adjustment for copy number relative to ploidy, tumour purity and cancer type (Spearman $\rho = -0.55$; $P = 0.0012$; **Fig. 5m**). Together, these findings link WGD-associated reductions in interferon-response and antigen-presentation programmes to immune-gene promoter hypermethylation, without identifying a corresponding excess of the genetic immune-escape alterations examined.

## Discussion

Understanding how a common genome-wide event shapes tumour evolution across distinct cancer lineages remains a fundamental question in cancer biology. Here we reveal an evolutionary pattern associated with whole-genome doubling (WGD) that extends beyond chromosome duplication. Tumours reach WGD through diverse genetic histories, yet subsequent chromosome-arm losses become substantially less dependent on cancer type. Beyond chromosomal evolution, established WGD+ tumours exhibit reduced interferon-response and HLA antigen-presentation programmes linked to epigenetic remodelling. Together, these findings position WGD as a recurrent evolutionary transition marked by altered lineage dependence of chromosomal evolution and epigenetically linked immune adaptation across cancers.

The preceding genetic background may shape which doubled clones succeed. *TP53* disruption is closely associated with WGD, while several drivers, including *KRAS*, *PIK3CA* and *PTEN*, show cancer-type-specific mutual exclusivity despite usually preceding doubling when present.[1,5] RAS–RAF–MAPK and DNA-repair protein assemblies show contrasting genomic associations.[17] These patterns raise the possibility that the additional fitness gained from WGD depends on the advantages already provided by early drivers. Doubling can buffer harmful chromosome losses and create opportunities for selection of increased oncogenic mutant-allele dosage,[3,18] while also increasing the demands of genome replication and cell maintenance. Newly formed tetraploid cells can lack replication factors in their first S phase, a defect alleviated experimentally by E2F1,[19] while mitochondrial DNA copy number increases in coordination with the nuclear genome.[20] The expansion of doubled clones may therefore depend on both the benefits of WGD and the capacity of cells to support a larger genome. Ovarian cancer provides an example of how this balance may differ across genetic backgrounds. WGD can arise repeatedly while only some doubled clones expand.[7] Although HRD-associated mutagenesis generally starts before WGD, WGD+ ovarian tumours show lower genomic HRD scores; Ewing and colleagues likewise found a negative HRD–WGD association using a different HRD measure.[11] WGD-high ovarian tumours are relatively enriched for BRCA2-like deletion-associated HRD and foldback-inversion backgrounds rather than BRCA1-like duplication-associated HRD.[7]

The genetic background inherited by a doubled clone may also shape the consequences of subsequent chromosome alterations. Chromosome-arm losses carry both growth benefits and survival costs: removing an arm changes the dosage of many genes, potentially weakening tumour suppressors alongside essential cellular functions.[21,22] The effect of a loss depends on both the genes affected and the number of copies that remain. WGD can soften some losses by preserving additional copies, while also altering chromosomal instability.[2,4] Earlier studies described broader aneuploidy landscapes in WGD+ tumours and distinct evolutionary patterns for chromosome gains and losses.[4,6] Our event-timing analysis identifies a further change across cancer types: cancer-type-explained variation in post-WGD arm-loss patterns falls by 61–75% across three cohorts, whereas driver mutations show no comparable overall reduction. Average loss patterns move closer across cancer types without reduced variation within them. Experimental chromosome-evolution screens show that tetraploid cells acquire more arm-level alterations yet retain tissue-specific fitness preferences;[23] fitness modelling likewise shows that the consequences of a change depend on the starting karyotype.[24] The reduced lineage dependence after WGD may therefore reflect changes in both which chromosome losses arise and which losses confer an advantage. By altering their costs and benefits, genome doubling may allow a broader range of chromosome losses to persist across different cancer lineages.

The evolutionary history established around WGD remains recognizable as tumours progress to metastasis. Primary and metastatic tumours share genomic ancestry but continue to diversify.[8,25] In matched primary–metastasis pairs, 89% of jointly timed events retained their placement before or after WGD as additional alterations accumulated. Primary-tumour WGD was also associated with subsequent metastasis, while WGD was more frequent in ovarian tumours exhibiting miliary peritoneal dissemination. The stronger WGD association of 8p loss of heterozygosity in metastases offers one genomic link to progression, consistent with chromosome-engineering experiments.[26] Other metastasis-associated alterations showed different relationships with WGD, indicating that the genomic changes accompanying metastatic progression are diverse.

Beyond these genomic alterations, metastatic outgrowth also depends on the ability of disseminated cells to survive immune surveillance. Chromosomal instability can promote metastasis through cytosolic DNA signalling,[27] while STING-related responses can favour metastatic growth during chronic inflammation or restrain the reactivation of dormant cells in other settings.[28,29] WGD+ tumours also show reduced T-cell-associated features and lower MHC class I antigen-presentation and interferon-response programmes within cancer cells. Patients with WGD+ tumours receiving immune checkpoint blockade also had shorter overall survival. Together, these findings identify genomic diversification and reduced immune recognition as distinct features associated with genome-doubled tumours. These observations suggest that changes in immune recognition may accompany the continued evolution and expansion of doubled clones during metastatic progression.

The reduced antigen-presentation programmes of established WGD+ tumours may reflect changes acquired as doubled clones expand. Immune surveillance can constrain hyperploid cells,[30] and experimental evolution can select aneuploid clones with attenuated antigen presentation.[31] Acute WGD can induce interferon-related transcription, whereas established WGD-high ovarian cancers exhibit reduced *STING1* expression despite ongoing chromosomal instability.[7,9,10] Foidart and colleagues established an experimental route from WGD to antigen-presentation silencing involving succinate accumulation, reduced KDM6 activity and increased

H3K27me3.[10] Our findings reveal another layer of epigenetic regulation associated with these immune programmes in human tumours. Broad hypomethylation within partially methylated domains, linked to cell division,[32] coexists with focal hypermethylation, a combination previously associated with immune-gene repression in genomically altered cancers.[33] WGD-associated hypermethylation is enriched at Polycomb-target CpG islands, which are particularly susceptible to cancer-associated DNA methylation.[34,35] This enrichment identifies a chromatin context in which local DNA hypermethylation preferentially accumulates. At the immune-gene level, WGD-associated hypermethylated regions are enriched for interferon and antigen-presentation programmes, and greater promoter hypermethylation is associated with lower expression. Experimental Polycomb repression of MHC class I genes and targeted epigenetic reactivation of *NLRC5* demonstrate that chromatin regulation can alter tumour-cell recognition.[36,37] Together, these findings suggest that promoter DNA methylation may contribute to maintaining reduced immune-gene expression in established WGD+ tumours, alongside histone-based regulation.

These evolutionary histories open two directions for future work: distinguishing how post-WGD chromosome losses arise from how they are selected, and testing whether WGD timing, preceding drivers and DNA-repair states improve clinical stratification. The mitotic vulnerabilities of doubled cells and epigenetic regulation of antigen presentation offer complementary therapeutic directions. Together, our findings position WGD as a recurrent evolutionary transition linking distinct early tumour histories to less lineage-dependent chromosome-arm-loss evolution, with much of this temporal ordering preserved during metastasis.

## Main References

## Figure legends

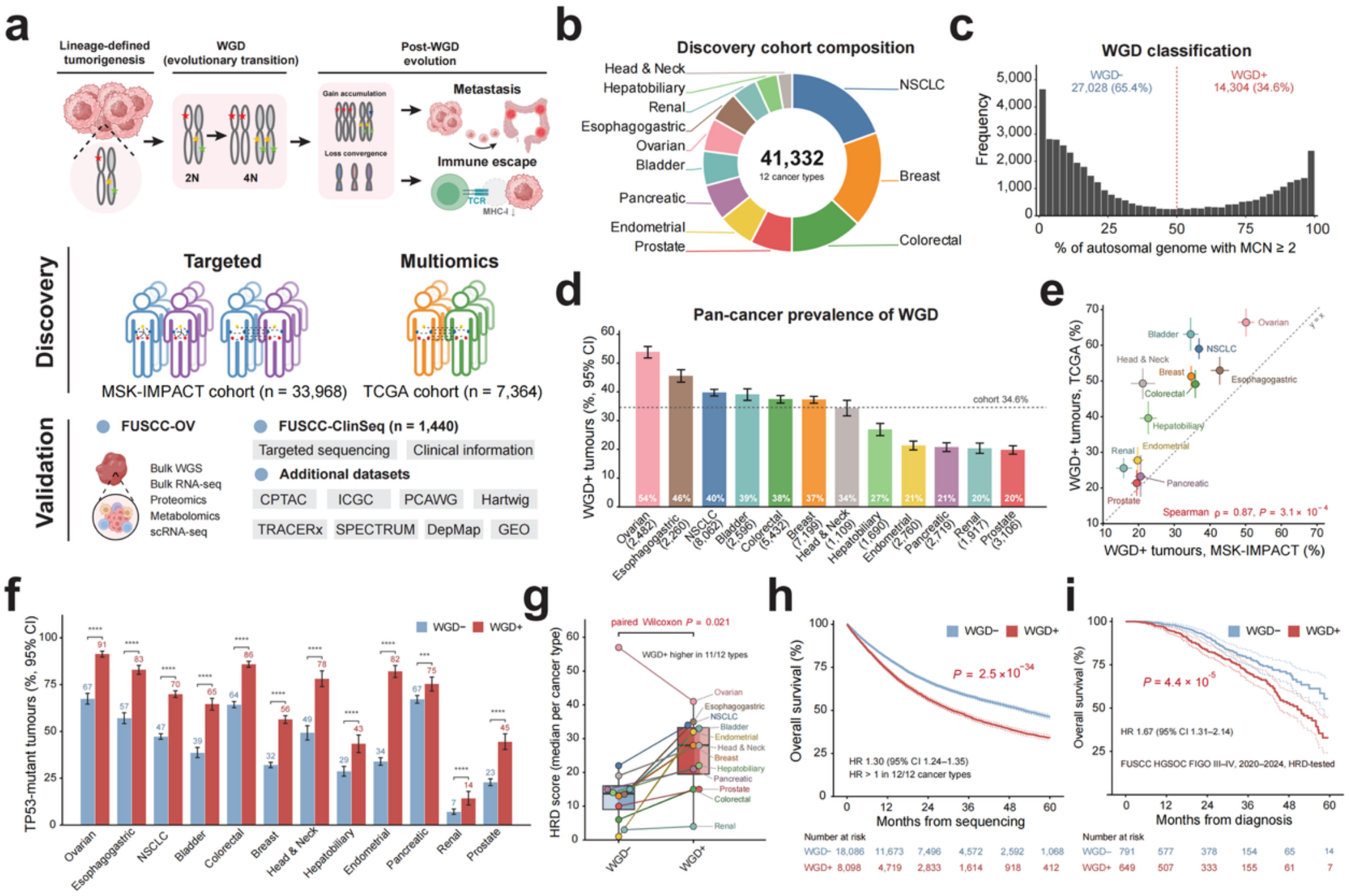

**Fig. 1 | Prevalence and correlates of whole-genome doubling across cancers. a**, Overview of the study design and cohorts. **b**, Cancer-type composition of the discovery cohort ($n = 41{,}332$ tumours). **c**, Distribution of the autosomal genome fraction with major copy number ≥2; dashed line indicates the WGD classification threshold. **d**, WGD prevalence across cancer types; dashed line indicates the overall prevalence. **e**, Comparison of WGD prevalence between MSK-IMPACT and TCGA (Spearman correlation). **f**, *TP53* mutation prevalence by WGD status ($n$ = 40,730 tumours). Brackets indicate two-sided Fisher's exact tests with Benjamini–Hochberg correction (***$q < 0.001$; ****$q < 0.0001$). Error bars in **d**–**f** indicate 95% patient-level cluster-bootstrap confidence intervals (CIs). **g**, HRD scores by WGD status in TCGA ($n$ = 6,056 tumours; two-sided paired Wilcoxon test across cancer-type medians). Lines connect cancer-type medians; boxes indicate medians and interquartile ranges, with whiskers extending to 1.5× the interquartile range. **h**, Overall survival from first sequencing in MSK-IMPACT ($n$ = 26,184 patients). **i**, Overall survival from diagnosis in FUSCC-ClinSeq high-grade serous

ovarian cancer ($n$ = 1,440 patients). Dotted lines in h,i indicate 95% CIs; $P$ values, Wald tests from the Cox models (Supplementary Table 2).

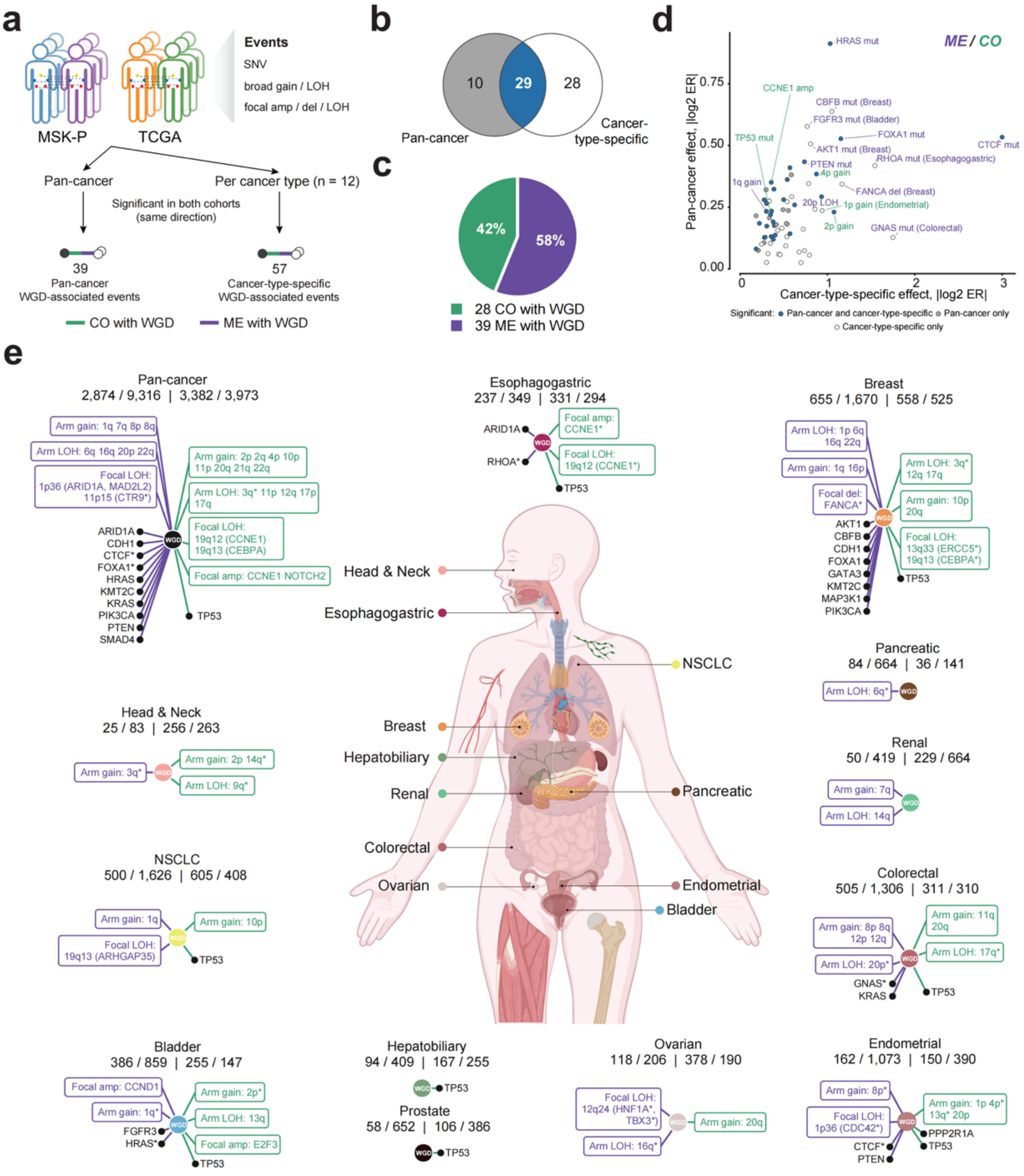

**Fig. 2 | Pan-cancer and lineage-specific genomic associations with WGD. a**, Framework for identifying WGD-associated genomic alterations in MSK-IMPACT and TCGA primary tumours, across all cancers and within individual cancer types. **b**, Overlap of alterations retained in pan-cancer and cancer-type-specific analyses. **c**, Distribution of retained alterations showing co-occurrence (CO) or mutual exclusivity (ME) with WGD. **d**, Pan-cancer and cancer-type-specific effect sizes (absolute $\log_2$ enrichment ratio (ER), averaged across cohorts). Point fill indicates retention scope, and label colour indicates association direction. **e**, Networks of

WGD-associated alterations across cancer types. Purple indicates ME and green indicates CO; black dots represent mutated genes, whereas boxes denote copy-number alterations. Focal loss-of-heterozygosity (LOH) regions are labelled by cytoband and representative genes. Numbers above each network indicate WGD+/WGD− tumours in MSK-IMPACT and TCGA. Asterisks denote alterations with false discovery rate (FDR) $\leq 0.10$ in one cohort and $P < 0.05$ in the other; unmarked alterations met FDR $\leq 0.10$ in both cohorts. ER, observed/expected number of WGD+ tumours with the alteration.

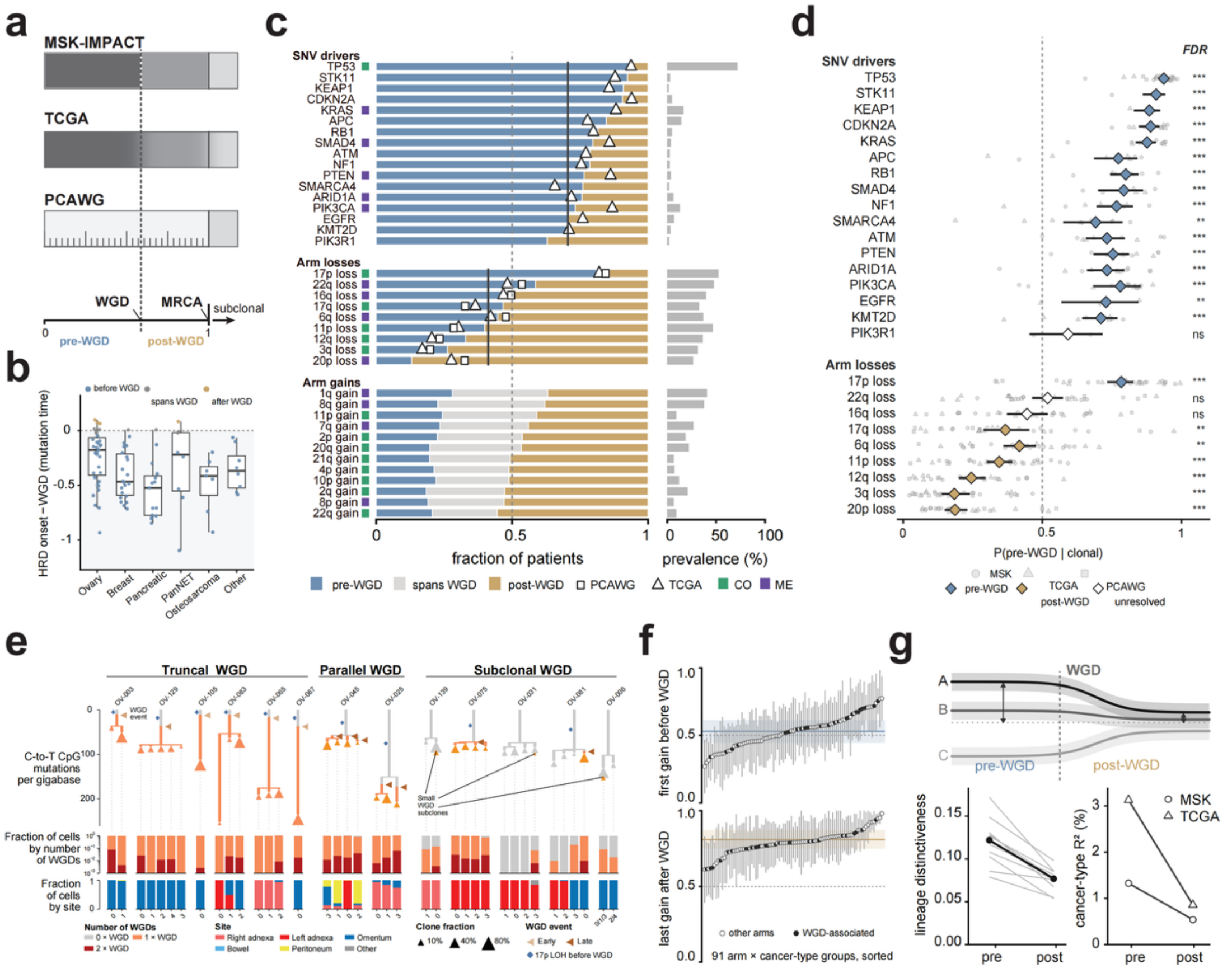

**Fig. 3 | Genomic event timing and reduced lineage dependence after WGD. a**, Event timing relative to WGD in MSK-IMPACT, TCGA and PCAWG. **b**, Onset of HRD-associated SBS3 relative to WGD in PCAWG (*n* = 95). Colours indicate whether bootstrap intervals precede, span or follow WGD. **c**, Timing of driver mutations, chromosome-arm losses and gains relative to WGD. Blue, pre-WGD; grey, spanning WGD; gold, post-WGD. Solid lines indicate mutation and arm-loss baselines; dashed lines, 50%. Green and purple symbols denote WGD co-occurrence and mutual exclusivity. **d**, Random-effects estimates of pre-WGD timing; bars, 95% confidence intervals (CIs); filled diamonds, CIs excluding 0.5; two and three asterisks indicate FDR < 0.01 and FDR < 0.001, respectively. **e**, Single-cell evolutionary histories from 13 high-grade serous ovarian cancers; diamonds indicate 17p loss of heterozygosity (LOH) before WGD, and bars show cell fractions by WGD count and sampling site. **f**, Probabilities that the first chromosome-arm gain preceded WGD and the last followed WGD across 91 arm–

cancer-type groups; bars, 95% CIs; line and band, random-effects mean ± between-group s.d. **g**, Lineage distinctiveness and cancer-type-explained variance (PERMANOVA $R^2$) before and after WGD in MSK-IMPACT and TCGA. HRD, homologous recombination deficiency.

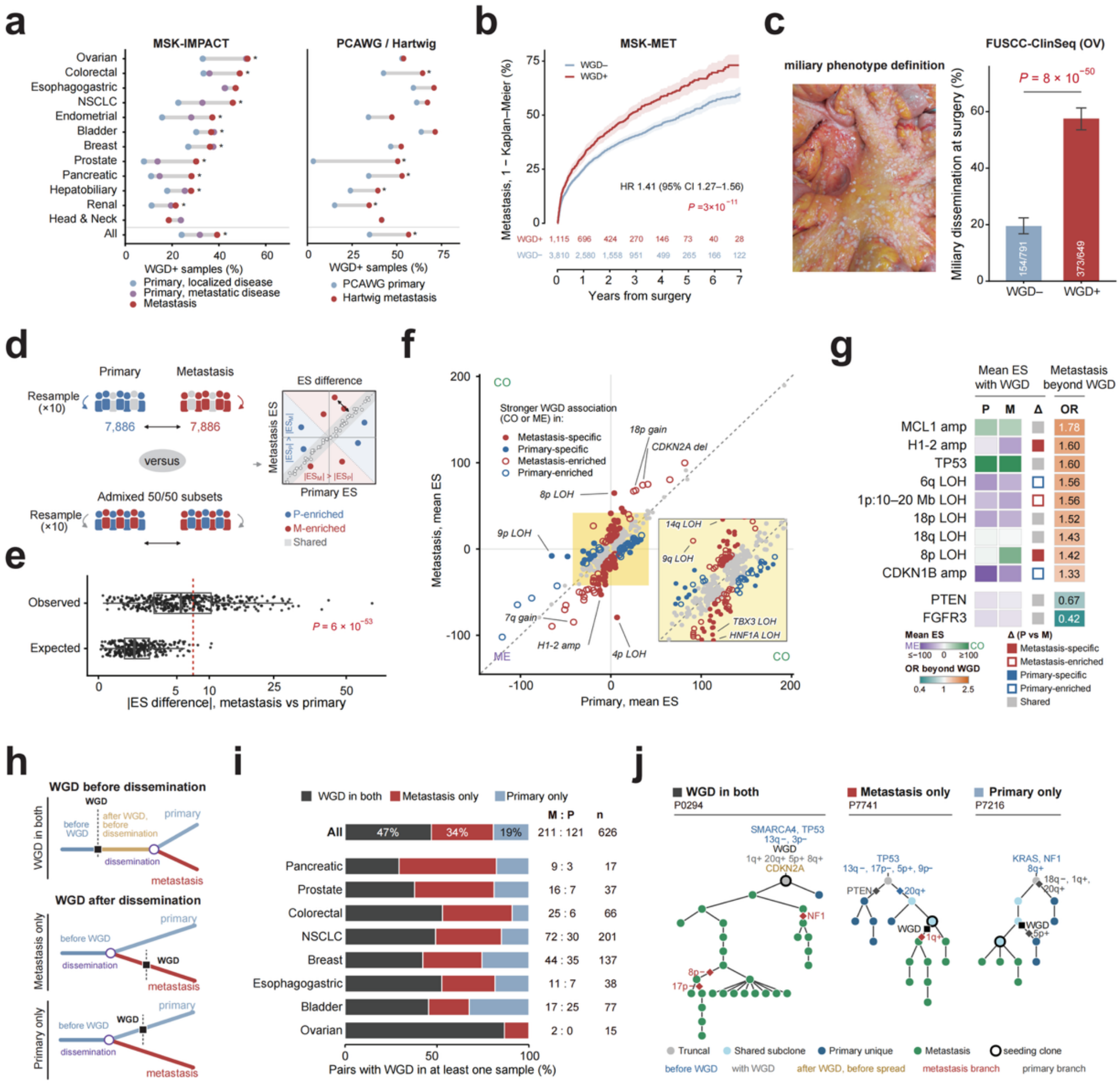

**Fig. 4 | Whole-genome doubling and metastatic progression. a**, WGD prevalence by disease state in MSK-IMPACT, and in PCAWG primaries versus Hartwig metastases. **b**, Metastasis after primary-tumour surgery in MSK-MET (n = 4,925), one minus the Kaplan–Meier estimate; shaded bands indicate 95% confidence intervals (CIs). **c**, Representative image of miliary peritoneal dissemination and its frequency by WGD status in FUSCC-ClinSeq (*n* = 1,440); error bars, 95% CIs. **d**, Cancer-type- and size-matched primary–metastasis comparisons and mixed-sample controls. **e**, Absolute differences in WGD-association effect size (ES) between primary and metastatic tumours versus controls; dashed line indicates the 99th percentile of controls. **f**, Primary versus metastatic ES across 395 loci; colours distinguish shared, site-specific and strengthened associations. **g**, WGD-association ES and adjusted odds ratios (ORs)

for metastatic status across selected alterations. **h**, Possible timing of WGD relative to metastatic dissemination. **i**, WGD detection patterns in 626 matched primary–metastasis pairs with WGD at one or both sites. **j**, Representative TRACERx–PEACE phylogenies showing WGD relative to metastatic seeding. *P* values were calculated using a cancer-type-stratified Cox model (**b**), Yates-corrected $\chi^2$ test (**c**) and two-sided Wilcoxon test (**e**).

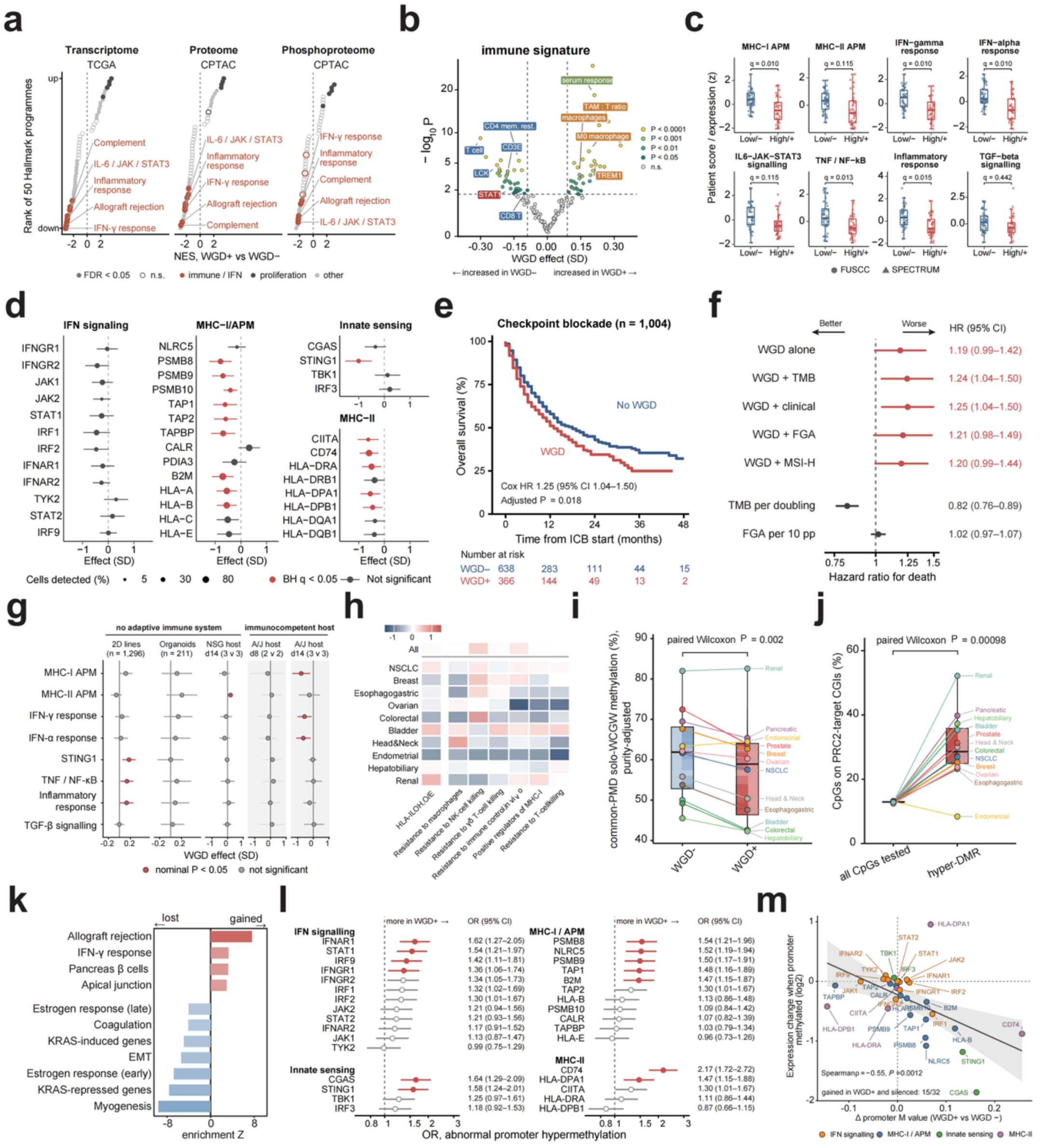

**Fig. 5 | Reduced antigen presentation and immune-associated methylation in WGD+ tumours.** **a**, WGD-associated Hallmark enrichment in TCGA transcriptomes and CPTAC proteomes and phosphoproteomes; filled points, FDR < 0.05. **b**, WGD effects on immune-compartment-normalized features in TCGA; dashed lines indicate the empirical FDR threshold. **c**, Single-cell RNA-seq-derived cancer-cell immune programme scores, summarized per patient (n = 76 patients with HGSOC) and compared by WGD status. **d**, Cohort-adjusted immune-gene expression differences in cancer cells from the same single-cell datasets; bars, 95% CIs; red, BH $q < 0.05$. **e**, Overall survival after immune checkpoint blockade ($n$ = 1,004);

Kaplan–Meier curves show 95% confidence bands. **f**, Hazard ratios from alternative Cox models; bars, 95% CIs. **g**, WGD effects on immune programmes in cell lines, organoids and isogenic mouse tumours; red, nominal $P < 0.05$. **h**, WGD associations with HLA-I loss of heterozygosity (LOH) and immune-escape gene sets across cancer types. **i**, PMD methylation by WGD status across cancer types. **j**, PRC2-target CpG-island enrichment among WGD-associated hypermethylated regions. Paired Wilcoxon tests in **i**,**j**. **k**, Hallmark enrichment among WGD-associated hyper- and hypomethylated regions (Stouffer Z). **l**, Odds ratios for immune-gene promoter hypermethylation; bars, 95% CIs; red, BH $q < 0.05$. **m**, Association between WGD-related promoter methylation differences and expression changes associated with promoter hypermethylation; points represent immune genes, and the line and band indicate the linear fit and 95% CI.

## Methods

### Study cohorts

The study was conducted in accordance with the Declaration of Helsinki and approved by the Ethics Committee of Fudan University Shanghai Cancer Center (FUSCC; protocols 2601-Exp365 and 2508-Exp538). Informed consent was waived for retrospective analyses of de-identified data, whereas written informed consent was obtained for prospective studies.

The discovery cohort comprised 41,332 tumours from 37,833 patients across 12 cancer types (**Supplementary Table 1**), including 33,968 tumours from 30,469 patients in the MSK-IMPACT 50K release[38] that passed FACETS quality control[39] and 7,364 primary tumours from The Cancer Genome Atlas (TCGA) with ASCAT allele-specific copy-number profiles obtained from the Genomic Data Commons.[40] Somatic mutations were obtained from MSK-IMPACT and TCGA MC3,[41] TCGA RNA-seq and DNA methylation data from public releases and UCSC Xena,[42] and clinical and metastatic outcomes from MSK-MET.[43] The immune checkpoint blockade analysis included 1,004 patients across seven cancer types selected from the Samstein cohort.[44]

FUSCC-ClinSeq comprised 1,440 patients with International Federation of Gynecology and Obstetrics (FIGO) stage III–IV high-grade serous ovarian cancer diagnosed between 2020 and 2024 and tested for homologous recombination deficiency, with targeted sequencing, complementary Affymetrix OncoScan copy-number profiling and surgical dissemination data.

FUSCC-OV contributed 132 bulk RNA-seq, 128 proteomic and 78 metabolomic tumour specimens, together with whole-genome sequencing data from 31 tumours for HRD–WGD timing analysis. Single-cell transcriptomic analyses included 149 pretreatment specimens from 76 patients (44 from FUSCC-OV and 32 from SPECTRUM).[7]

Independent genomic validation included 1,157 non-TCGA tumour genomes from 1,069 donors in the International Cancer Genome Consortium (ICGC)/Pan-Cancer Analysis of Whole Genomes (PCAWG),[45] of which 1,126 (1,038 donors) had somatic mutation data. Evolutionary analyses incorporated published gain-timing posteriors from 492 PCAWG tumours,[6] whole-genome data from 1,877 WGD+ Hartwig metastatic carcinomas,[46] multiregion sequencing data from TRACERx[47] and metastatic phylogenies from TRACERx–PEACE.[8] Single-cell genomic profiles and clonal phylogenies were obtained from SPECTRUM,[7] and single-cell ATAC-seq data from Sundaram and colleagues through the Genomic Data Commons.[48]

Additional molecular datasets included proteomic and phosphoproteomic data from the Clinical Proteomic Tumor Analysis Consortium (CPTAC), DNA methylation data from ICGC PACA-AU and the ovarian cancer cohort GSE292140, which combines the AOCS (GSE65820)[49] and MOCOG (GSE211686)[50] series with 24 additional EPIC arrays.[51] Experimental-model analyses used DepMap cancer cell lines, NextGen cancer organoid models,[52] published isogenic mouse mammary tumour single-cell RNA-seq and ATAC-seq datasets (GSE274140 and GSE274160),[10] and chromatin-compartment annotations from post-WGD *TP53*-null retinal pigment epithelial tumour models reported by Lambuta and colleagues.[9]

### Sample collection and molecular profiling

***Sample collection.*** Fresh tumour specimens were collected from patients enrolled in FUSCC-OV and processed for whole-genome sequencing, bulk and single-cell RNA sequencing, proteomics and metabolomics (Shanghai Applied Protein Technology Co., Ltd., Shanghai, China). Fresh tissue was used predominantly for single-cell RNA sequencing, whereas specimens for the other assays were generally frozen after collection and stored in liquid nitrogen. For specimens embedded in optimal cutting temperature (OCT) compound, tissue free of embedding medium was excised for DNA and RNA extraction.

***Whole-genome sequencing.*** Tumour whole-genome sequencing data were analysed using nf-core/oncoanalyser v2.3.0 in tumour-only mode with the HMF GRCh38 reference genome. PURPLE v4.3[46] integrated B-allele frequencies from AMBER v4.2, read-depth ratios from COBALT v2.2 and structural-variant breakpoints to estimate tumour purity, ploidy and allele-specific copy number. WGD status was obtained directly from the PURPLE wholeGenomeDuplication output.

***Bulk RNA sequencing.*** Total RNA was extracted using TRIzol and assessed for quantity and integrity. Polyadenylated RNA was enriched using oligo(dT) beads, and libraries were prepared using the BGI Optimal Dual-mode mRNA Library Prep Kit, followed by circularization, DNA nanoball generation and paired-end DNBSEQ sequencing. Raw reads were filtered using SOAPnuke v2.3.2[53] to remove adaptor contamination, reads containing more than 5% ambiguous bases and reads in which more than 40% of bases had Phred quality scores below 20. Filtered reads were aligned to the reference transcriptome using Bowtie2 v2.4.5[54] and quantified using RSEM v1.3.1.[55]

***Proteomics.*** Tumour tissue was homogenized and lysed in 5% sodium deoxycholate and 100 mM Tris-HCl (pH 8.5). Proteins were reduced with dithiothreitol, alkylated with iodoacetamide and digested with trypsin (enzyme-to-protein ratio, 1:50) for 15–18 h at 37 °C. Desalted peptides were analysed by data-independent acquisition mass spectrometry using an Orbitrap Astral mass spectrometer coupled to a Vanquish Neo liquid chromatography system. MS1 scans covered m/z 380–980 at a resolution of 240,000 (at m/z 200), with MS2 acquisition using 299 isolation windows of 2 m/z. A pooled quality-control sample (MixQC) was prepared by combining aliquots from all study samples and injected after every 12 study-sample injections.

Data were processed using DIA-NN v1.8.1,[56] allowing one missed tryptic cleavage, fixed cysteine carbamidomethylation and variable methionine oxidation and protein N-terminal acetylation. Protein identification was controlled at a false discovery rate (FDR) of 1%.

***Metabolomics.*** Tumour tissue was homogenized in water and extracted with methanol–acetonitrile (1:1, v/v). Following centrifugation, supernatants were dried and reconstituted in acetonitrile–water (1:1, v/v). Metabolites were separated by hydrophilic interaction chromatography using a BEH Amide column (2.1 × 100 mm, 1.7 μm) and analysed in positive and negative electrospray ionization modes using an Orbitrap Exploris 480 mass spectrometer. A pooled quality-control sample (MixQC) was prepared by combining aliquots from all study samples and injected after every 12 study-sample injections. Raw data were processed using ProteoWizard[57] and XCMS[58] to generate peak-area matrices. Metabolites were putatively annotated using accurate mass (mass-error tolerance, 25 ppm) and MS/MS spectral matching against an in-house reference library.

***Single-cell RNA sequencing.*** Fresh tumour tissue was minced and dissociated enzymatically and mechanically at 37 °C. Cell suspensions were filtered through 70-μm and 30-μm strainers, followed by erythrocyte lysis. Cell viability was assessed by acridine orange/propidium iodide staining, and suspensions with viability below 80% underwent dead-cell removal (Miltenyi Biotec). Single-cell 3′ gene-expression libraries were prepared using the Chromium X platform and Chromium Next GEM Single Cell 3′ Reagent Kit v3 (10x Genomics),[59] followed by paired-end sequencing. Gene–cell count matrices were generated using Cell Ranger.

**WGD classification**

Whole-genome doubling (WGD) in the MSK-IMPACT and TCGA discovery cohorts was defined as a major-allele copy number of at least 2 across more than 50% of the autosomal genome.[1] Major-allele copy number was calculated as total minus minor-allele copy number, with genomic fractions weighted by segment length. For MSK-IMPACT, the published FACETS-based WGD calls,[38,39] were used for all statistical analyses. For TCGA, one allele-specific copy-number profile per tumour was selected in the order ASCAT3, ASCAT2 and AscatNGS,[40] and WGD was assigned using the same length-weighted criterion.

For FUSCC-OV, WGD status was obtained directly from the wholeGenomeDuplication field of PURPLE v4.3,[46] which identifies WGD when more than ten autosomes have major-allele copy number greater than 1.5. For FUSCC-ClinSeq, WGD status and homologous recombination deficiency (HRD) scores were determined using Affymetrix OncoScan arrays. Other public datasets retained their source-specific WGD annotations unless otherwise specified in the corresponding analyses.

**WGD prevalence and clinical correlates**

WGD prevalence was calculated by cancer type in the pooled MSK-IMPACT–TCGA discovery cohort. Ninety-five per cent confidence intervals (CIs) were estimated using 5,000 patient-level bootstrap resamples, retaining all tumours from each patient. Agreement in prevalence across the 12 cancer types was assessed by Spearman correlation between cohorts.

TP53 mutations were identified from MSK-IMPACT and TCGA MC3 v0.2.8,[41] retaining non-silent variants. A total of 40,730 tumours had matched mutation and copy-number data. Associations with WGD were tested using Fisher's exact test within each cancer type, with Benjamini–Hochberg correction for multiple testing. Pooled odds ratios were estimated using the Cochran–Mantel–Haenszel method stratified by cancer type. Published TCGA homologous recombination deficiency (HRD) scores,[60] calculated as the sum of loss of heterozygosity, telomeric allelic imbalance and large-scale state transitions, were available for 6,056 tumours. Cancer-type median HRD scores were compared between WGD groups using the paired Wilcoxon signed-rank test.

For MSK-IMPACT, overall survival was measured from first sequencing, with WGD and TP53 status assigned from each patient's first eligible tumour (26,184 patients; 10,154 deaths). Cox proportional-hazards models were stratified by cancer type and disease state at sequencing, with additional adjustment for TP53 where indicated. For FUSCC-ClinSeq, survival was measured from the date of pathological diagnosis (1,440 patients; 265 deaths). Multivariable Cox models adjusted for age, FIGO stage IV, clinical HRD, log-transformed CA-125, complete macroscopic resection and neoadjuvant chemotherapy. Analyses used the R package survival v3.8-11. Kaplan–Meier curves were unadjusted, and Cox-model P values were derived from Wald tests (**Supplementary Table 2**).

**WGD-associated genomic alterations**

Genomic associations with WGD were assessed in primary tumours from MSK-IMPACT (n = 12,190; localized disease) and TCGA (n = 7,355, including 6,753 with MC3 mutation calls), excluding neuroendocrine tumours. Independent validation included 1,157 non-TCGA ICGC/PCAWG tumour genomes from 1,069 donors (1,100 primary tumours, 45 metastases, 11 recurrences and one tumour-derived cell line), of which 1,126 (1,038 donors) had mutation data; genomes from the same donor were analysed as separate samples. Gene-level events required at least one non-synonymous mutation. MC3 calls were restricted to PASS, wga or native_wga_mix and deduplicated to one call per tumour and genomic position.[41] PCAWG variants were restricted to coding regions and 2-bp flanks of tested genes (GENCODE v19).[61]

Copy-number alterations were classified across 39 autosomal chromosome arms, excluding acrocentric short arms. Arm-level gains required total copy number above a WGD- and ploidy-adjusted baseline[25] (2 for WGD−, 4 for WGD+ with ploidy <6 and 8 otherwise) across at least 80% of the covered arm length. Arm-level loss of heterozygosity (LOH) required minor-allele copy number zero across at least 80% of the covered arm. Focal alterations spanned less than half an arm; amplification required at least twice the arm-modal copy number and two copies above baseline, whereas deletion required zero or no more than one-quarter of the arm-modal copy number. Focal LOH, defined by minor-allele copy number zero without arm-level LOH, was assessed in 10-Mb loci. TCGA high-level focal amplifications and deletions were defined using GISTIC calls (+2 and −2, respectively).[62]

MSK-IMPACT analyses were stratified by detailed cancer type, sequencing panel and microsatellite status, and TCGA analyses by project and microsatellite status. Microsatellite instability was defined using released MSK annotations or TCGA MSIsensor scores ⩾10.[63,64] Hypermutated tumours (>10-fold the stratum median mutation burden) were assigned to separate mutation strata. Cancer-type-specific analyses required at least 80 tumours, including ten in each WGD group.

We adapted a burden-adjusted simulation framework developed for pairwise co-mutation analysis[65] to assess associations between WGD and genomic alterations. The null model fixed WGD labels and redistributed each event's observed carriers within strata, with sampling probabilities proportional to tumour burden for that event class. Copy-number and mutation analyses used 10,000 and 2,000 simulations, respectively. For event *e*, enrichment was $\mathrm{ER}_e = O_e/E_e$, where $O_e$ was the observed number of WGD+ carriers and $E_e$ the simulated expectation. Normalized effect sizes (nES) were obtained by subtracting the mean absolute null effect from the magnitude of $O_e - E_e$, retaining its sign and truncating negative magnitudes at zero. Empirical false discovery rates (FDRs) were estimated within event-frequency bins (⩽ 2%, >2 – 5%, >5 – 10% and >10%) to account for frequency-dependent variation in null effects. At each absolute nES threshold, FDR was estimated as the mean number of simulated events reaching the threshold divided by the corresponding number of observed events. Associations with FDR ⩽0.10 were classified as co-occurring (ER >1) or mutually exclusive (ER <1) with WGD.

Null-model calibration used five WGD-independent simulated cohorts per event class and discovery cohort, with silent mutations as negative controls. Arm-level copy-number loss was excluded from association testing owing to poor null calibration but retained for timing analyses. Discovery associations required concordant directions across cohorts and either FDR ≤0.10 in both cohorts or FDR ≤0.10 in one cohort and nominal $P$ <0.05 in the other. Independent validation significance was assessed in ICGC/PCAWG at FDR ≤0.10, without affecting discovery retention; directional concordance was evaluated separately. Effect-magnitude comparisons used mean $|\log_2 \text{ER}|$ across discovery cohorts, capped at 3 for visualization; cancer-type comparisons used the most significant eligible cancer type (**Supplementary Table 3**).

Protein-assembly analysis used the NeST hierarchy of 131 multiprotein assemblies, frozen from its public release[12,13]; node coordinates and hierarchy edges were taken verbatim from the deposited network. Assemblies with more than 100 genes, and those with fewer than five genes represented on the sequencing panel, were not tested, leaving 87 of the 131. For each assembly the per-sample statistic was the fraction of its panel genes carrying a non-synonymous mutation. Per-gene effect sizes against WGD were taken from the same null model as the locus-level analysis and rank-transformed using van der Waerden normal scores. Significance combined a competitive test, permuting gene labels within strata of cancer type and mutation count outside the assembly, with a self-contained one-sample test of the member scores; assemblies were retained at competitive FDR ⩽0.10 and self-contained P <0.05 with concordant direction in MSK-IMPACT and TCGA. Burden was matched outside the assembly rather than over the whole panel because matching on the total count nearly fixes a large assembly's statistic: under total-burden matching none of the eight assemblies with at least 50 genes reached significance, against 62% of those with 20 to 49 genes. Sensitivity analyses repeated the test on untransformed effect sizes and after removing TP53 from the gene pool. Protein-association edges were retrieved from STRING v12.0.[66]

**Evolutionary timing of genomic events**

Mutation and chromosome-arm-loss timing was assessed in 10,914 WGD+ MSK-IMPACT specimens from 10,071 patients, 2,183 TCGA tumours with ABSOLUTE and MC3 profiles, and 417 microsatellite-stable PCAWG tumours with a single WGD event and sufficient sequencing coverage. TCGA mutation timing used ABSOLUTE[67] rather than the ASCAT profiles used for WGD classification. External analyses included TRACERx multiregion lung tumours[47] and 1,877 WGD+ Hartwig metastatic carcinomas.[46] Chromosome-gain timing used published joint posterior draws from 492 PCAWG tumours and corresponding Hartwig subsets.[6]

MSK-IMPACT timing used released FACETS mutant multiplicity and clonality. On allele-specific states 2+2, 3+3 and 2+0, clonal mutations with at least two mutant copies were assigned pre-WGD and those with one copy post-WGD; other states were not timed. In TCGA and PCAWG, read counts, purity and allele-specific copy number informed posterior probabilities of clonal mutant multiplicity and subclonality. Expected variant allele fractions were

$$f_m = \frac{\rho m}{\rho C + 2(1-\rho)}, \qquad f_\phi = \frac{\rho \phi}{\rho C + 2(1-\rho)},$$

where $\rho$ denotes purity, $C$ total copy number, $m$ clonal mutant copies and $\phi$ the subclonal cancer-cell fraction. Equal prior weight was assigned to clonal and subclonal classes; $m = 1, \dots, M$ and $\phi = 0.05, 0.15, \dots, 0.95$ were uniform within their respective classes.[5] Binomial read likelihoods yielded $P_{\text{pre}}$ (multiplicity ≥2), $P_{\text{post}}$ (multiplicity 1) and $P_{\text{sub}}$; for segments with major-allele copy number <2, pre- and post-WGD timing probabilities were set to zero. TRACERx used published early- and late-clonal assignments.

Chromosome-arm losses affecting at least 80% of an arm were classified as pre-WGD for allele-specific state 2+0 and post-WGD for 2+1; other states remained unresolved. TRACERx regional calls were aggregated by majority, with LOH shared between WGD− and WGD+ regions supporting pre-WGD timing and WGD+-restricted LOH supporting post-WGD timing. Each patient contributed one record per event. Pre-WGD fractions were $n_{\text{pre}}/(n_{\text{pre}} + n_{\text{post}})$ in MSK-IMPACT and $\sum P_{\text{pre}}/\sum(P_{\text{pre}} + P_{\text{post}})$ in TCGA/PCAWG. Within each platform and each of the 12 cancer types (at least five timed patients), per-patient pre-WGD probabilities were summed into effective counts, logit-transformed with a half-count continuity correction and shrunk towards the platform mean by empirical Bayes; the cancer-type estimates were then pooled by DerSimonian–Laird random-effects meta-analysis, with heterogeneity summarized as $I^2$ across cancer types and across platforms. Driver analyses used OncoKB oncogenic or likely oncogenic variants; thresholds were 100 timed MSK patients or 20 TCGA/PCAWG tumours.[68] Two-sided tests against 0.5 and Benjamini–Hochberg correction were applied (**Supplementary Table 4**).

Chromosome-gain timing was inferred from published joint posterior draws[6] estimating the probabilities that the first arm-level gain preceded WGD and the last gain followed WGD. Estimates were pooled across arm-by-cancer-type strata with at least ten tumours using restricted maximum-likelihood random-effects meta-analysis (metafor v5.0.1; stratum standard errors floored at 0.02), separately for PCAWG and Hartwig.

HRD timing was inferred using HRDTimer,[69] which integrates mutation multiplicity with SBS1 and SBS3 signature probabilities. Eligible WGD+ PCAWG tumours required at least 50 SNVs, an SBS3 contribution ⩾10% and detectable contributions from both signatures. HRD onset was estimated using 200 mutation bootstraps and compared with WGD timing inferred using MutationTimeR v1.0.2.[5] In FUSCC-OV, HRD was defined by an ovaHRDscar score ⩾ 54,[70] calculated from PURPLE segments ⩾100 kb. Thirty-one WGD+ tumours with estimable HRD and WGD timings were analysed on the SBS1 molecular time scale. The proportion with HRD onset preceding WGD was assessed using a two-sided binomial test.

**Genomic convergence and single-cell reconstruction**

Convergence analysis included mutations and arm losses timed before or after WGD and observed in at least 20 tumours in each epoch. Tumours required at least three events per epoch and cancer types at least 100 tumours (MSK-IMPACT: 7,313 tumours, ten types; TCGA: 1,298 tumours, six types). Across 200 draws, three events per tumour and epoch were sampled without replacement, generating normalized event vectors. Cancer-type explained variance ($R^2$, equivalent to Euclidean PERMANOVA) and lineage distinctiveness ($D_g$) were calculated from the distances between cancer-type and pan-cancer event centroids. Lineage distinctiveness was defined as the distance from each cancer-type centroid to the pan-cancer centroid divided by within-type dispersion. The primary contrast was $R^2_{\text{post}} - R^2_{\text{pre}}$; significance was assessed by within-tumour epoch-label permutation (2,000 permutations; 500 for arm-loss-only analyses), preserving event counts. The same test was applied to Hartwig metastatic whole genomes using arm losses only, since the released timing data contain allele-specific segments but no mutation calls; cancer types with at least 100 tumours gave three types and 712 tumours, and lowering this threshold to 50 gave six types and 918 tumours. Convergence was additionally decomposed into the distance between cancer-type and pan-cancer centroids and the dispersion within cancer types, and summarized by leave-one-out nearest-centroid classification of cancer type (**Supplementary Table 5**).

Leave-one-out nearest-centroid classification used the same rarefied event matrices as the primary convergence analysis. Per-event lineage dependence was quantified by each event's contribution to the cancer-type-by-event $\chi^2$ statistic, averaged across sampling draws and normalized by its sampling frequency. Pre- and post-WGD values were compared separately for chromosome-arm losses and driver mutations using paired two-sided Wilcoxon signed-rank and sign tests, with median per-event relative changes reported (**Supplementary Table 6**).

Single-cell clone trees and allele-specific copy-number profiles from SPECTRUM were used[7] to time chromosome-arm losses relative to WGD in 25 high-grade serous ovarian cancer trees. Clone-level analyses required at least five once-doubled cells and a dominant allele-specific copy-number state covering at least 60% of an arm. Losses were classified as pre-WGD for state 2+0 and post-WGD for 2+1. Concordance with bulk ovarian timing was assessed by Spearman correlation. Additional ovarian, breast, lung, renal and colorectal single-cell datasets were analysed using published clone assignments or phylogenies reconstructed with MEDICC2 v1.1.2.[71] Alterations shared by undoubled and doubled clones were classified as pre-WGD, whereas those restricted to doubled clones were classified as post-WGD relative to clone ploidy.

**Metastatic progression and evolution**

FACETS-quality-controlled MSK-IMPACT specimens were classified as primary tumours from patients with localized or metastatic disease, or as metastases. Microsatellite-unstable and neuroendocrine tumours were excluded. WGD prevalence trends were assessed using cancer-type-specific Cochran–Armitage tests, and metastatic sites were compared with primary tumours using logistic regression adjusted for cancer type. WGD prevalence was also compared between PCAWG primary tumours and Hartwig metastases using harmonized published calls. Subsequent metastasis was analysed in 4,925 MSK-MET patients without metastasis at primary-tumour surgery, using one minus the Kaplan–Meier estimate and Cox regression stratified by cancer type. Miliary dissemination in the 1,440 FUSCC-ClinSeq patients was compared between WGD groups using the $\chi^2$ test.

To compare WGD-associated genomic alterations between primary and metastatic tumours, equal numbers of specimens were resampled within each of 37 detailed cancer types (7,886 per group; ten resamples), alongside matched-proportion mixed-site controls. The WGD-association analysis was repeated in each cohort. Loci significant in at least one resample were retained, and differences in association effect sizes were compared against the mixed-control distribution. Differences exceeding the 99th percentile were considered significant; site specificity additionally required significance in at least six of ten resamples at one site and no more than two at the other.

Metastasis-associated alterations were independently identified in 9,561 MSK-MET patients using elastic-net logistic regression (scikit-learn v1.7.2; mixing parameter, 0.5) with stability selection across 100 outcome-stratified half-samples. The penalty was chosen to retain approximately 15% of features, and events selected in at least 70% of resamples were retained. Odds ratios were estimated using separate unpenalized logistic models adjusted for WGD, cancer type, sequencing panel and tumour purity (n = 9,556 with complete purity data).

Paired analyses included 1,036 MSK-IMPACT patients with FACETS-quality-controlled primary and metastatic samples (purity ≥0.2). Each site was classified as WGD+ if any available specimen was positive; representative specimens were selected by prioritizing WGD positivity and then the number of timed events. Among discordant pairs, WGD detection was compared between sites using logistic regression adjusted for differences in tumour purity. Event timing concordance in pairs with WGD at both sites was assessed using percentage agreement and Cohen's κ, and private post-WGD events were compared using the paired Wilcoxon test. Published TRACERx–PEACE phylogenies[8,72] were used to reconstruct the ordering of genomic alterations, WGD and metastatic seeding, leaving events on unresolved branches unassigned.

### Molecular and immune analyses

***Transcriptomic and proteomic pathway analysis.*** TCGA transcriptomes (n = 8,951; 32 cancer types), CPTAC proteomes (n = 976) and phosphoproteomes (n = 865), and FUSCC-OV RNA-seq (n = 132) and proteomic (n = 128) data were analysed for WGD-associated differences using limma v3.62.1.[73] TCGA RNA models adjusted for cancer type, whereas CPTAC and FUSCC-OV analyses were performed within cancer type. FUSCC-OV RNA counts were TMM-normalized and analysed using limma-voom,[74] whereas protein intensities were $\log_2$-transformed, median-normalized and analysed with robust empirical Bayes moderation. TCGA genes required $\log_2$ expression >1 in at least 20% of tumours; CPTAC proteins and phosphosites required ≥50% coverage and at least three measurements per WGD group. Signed statistics were combined across at least three cancer types using Stouffer's method. Enrichment of Hallmark gene sets used fgsea v1.32.2 with Benjamini–Hochberg correction within each analysis (**Supplementary Table 8**).[75,76]

***Metabolomic analysis.*** FUSCC-OV metabolomic data were analysed separately by ionization mode. Features required pooled quality-control relative standard deviation ≤30% and detection in ≥80% of specimens. Intensities were median-normalized and log-transformed without missing-value imputation. A total of 726 compounds meeting KEGG annotation, spectral-match (≥0.70) and detection criteria were analysed using limma, followed by enrichment analysis of 48 KEGG metabolic pathways with Benjamini–Hochberg correction.

***Immune composition and malignant-cell programmes.*** Immune composition was assessed using 162 published signatures in 5,847 TCGA tumours.[77] Within-type regressions adjusted for proliferation, with effects pooled by random-effects meta-analysis and empirical false discovery rates (FDRs) estimated from 1,000 within-type WGD-label permutations. Malignant-cell immune programmes were analysed in 76 patients from FUSCC-OV and SPECTRUM after doublet removal and cell annotation. SPECTRUM[7] tumours were classified as WGD-high when more than half of sequenced cancer cells had undergone WGD; FUSCC-OV used tumour-level WGD calls. Programme scores and gene expression were aggregated at patient level and compared with cohort adjustment and Benjamini–Hochberg correction across eight programmes and 38 genes (**Supplementary Table 9**).

***Experimental models.*** Immune-related transcriptional programmes and *STING1* expression were assessed in 1,296 DepMap cell lines and 211 NextGen organoids[52] using released WGD annotations, adjusting for cancer lineage and proliferation. Published isogenic mouse mammary tumour scRNA-seq data (GSE274140) were analysed at tumour level across immunocompetent A/J and immunodeficient NSG hosts, with two or three mice per WGD group per setting.[10] Programme scores were compared using two-sided t tests, and host-by-WGD interactions were assessed at day 14. Corresponding chromatin accessibility data (GSE274160) provided descriptive comparisons.[10]

***Immune checkpoint blockade.*** Overall survival following immune checkpoint blockade was analysed in 1,004 MSK-IMPACT patients (517 deaths) from the Samstein cohort.[44] Follow-up began at treatment initiation. Cox models were stratified by cancer type and sequentially adjusted for log-transformed tumour mutational burden, age, sex, specimen site, drug class and either fraction of genome altered (absolute $\log_2$ copy ratio >0.2) or microsatellite-instability status. Kaplan–Meier curves were unadjusted.

***Immune-related genomic alterations.*** HLA class I-region loss of heterozygosity (LOH) was assessed using ASCAT profiles from 6,880 TCGA tumours, requiring minor-allele copy number zero across at least 80% of chr6:29.9–31.5 Mb. Regional LOH enrichment was evaluated relative to genome-wide LOH burden using binomial regression with a burden offset. Six immune-escape gene sets (49–251 genes) were curated from 67 experimental arms of 31 published CRISPR screens, informed by previous integrative analyses of immune escape.[78] Genes required knockout-associated immune escape in at least two studies, ≥60% directional agreement and FDR ≤0.10. Nonsense and essential-splice mutations in these genes were assessed in 6,266 TCGA tumours after excluding hypermutated samples (>1,000 coding SNVs; **Supplementary Table 10**).

***DNA methylation and chromatin analysis.*** DNA methylation was analysed in 5,139 tumours across 12 cancer types from TCGA, CPTAC, ICGC PACA-AU and the ovarian cancer cohort GSE292140 (**Supplementary Table 11**).[51] Cohort-specific WGD and tumour-purity estimates were used. Following established probe-quality masking and SNP-probe exclusion,[79] 346,015 shared autosomal probes were retained. Global methylation was measured as the median β value at 4,197 solo-WCGW CpGs within common partially methylated domains, using published reference annotations.[32] Within-cancer-type models adjusted for tumour purity and cohort block, with WGD contrasts compared across cancer types using paired Wilcoxon tests.

Differential methylation was assessed in 5,136 tumours using limma on M values, adjusting for WGD, tumour purity and cohort block. Probes required ≥95% detection within cancer type, with residual missing values imputed by ten nearest neighbours. Differentially methylated positions were identified at FDR <0.05, and regions were called using DMRcate v3.2.1 ($\lambda$ = 1,000; C = 2; ≥2 CpGs).[80] PRC2-target CpG islands were defined using the reference annotations of Hetzel et al., derived from bivalent and Polycomb-repressed chromatin states in HUES64 embryonic stem cells (Roadmap E016).[35] Their enrichment among WGD-associated hypermethylated regions was compared across cancer types using paired Wilcoxon tests. Hallmark enrichment of hyper- and hypomethylated regions was assessed using missMethyl v1.40.0,[81,82] with the tested-probe background and one-sided probabilities combined across cancer types using Stouffer's method. Published post-WGD *TP53*-null RPE chromatin data were analysed[9] for PRC2-target enrichment in regions shifting towards inactive compartments using Fisher's exact tests (**Supplementary Table 12**).

***Immune-gene promoter methylation.*** Promoter methylation was assessed for 38 prespecified immune genes spanning interferon signalling, antigen presentation, cGAS–STING and MHC-II regulation. Promoters were defined as −1,500 to +500 bp around transcription start sites; 32 genes had at least two quality-controlled probes. Hypermethylation was defined relative to the within-cancer-type 95th percentile of WGD− tumours and compared using cancer-type-stratified Cochran–Mantel–Haenszel analysis. Matched TCGA methylation and expression data (n = 4,116) were analysed by regression of gene expression on promoter hypermethylation, ploidy-relative copy number, tumour purity and cancer type. WGD-associated promoter methylation effects were pooled by random-effects meta-analysis and correlated with hypermethylation-associated expression differences using Spearman correlation.

***Single-cell chromatin accessibility.*** Cancer-cell scATAC-seq profiles[48] were assigned tumour-level WGD calls from ASCAT. Gene activity was calculated using Signac v1.17.1[83] over gene bodies and 2 kb upstream. Exploratory WGD comparisons used Seurat v5.5.1[84] cell-level Wilcoxon tests in breast (ten WGD+, five WGD− tumours) and colorectal cancers (five WGD+, seven WGD− tumours).

**Statistical analysis**

Statistical analyses were performed using R v4.4.3 and Python v3.10.20. Unless otherwise specified, all statistical tests were two-sided. Multiple-testing correction was performed using the Benjamini–Hochberg procedure or empirical false discovery rate estimation, as appropriate. Statistical methods and significance thresholds are detailed in the corresponding Methods sections and figure legends.

**Code availability**

Code implementing the computational methods developed in this study is available via GitHub at https://github.com/pigudog/wgdlife. Additional scripts for data processing, statistical analyses and reproduction of the main figures are available at https://github.com/pigudog/wgdlife_analysis.

**Data availability**

The sequencing data generated in this study are being deposited in the Genome Sequence Archive for Human (GSA-Human) under submission ID subHRA032831. The accession number will be provided upon completion of data processing. The curated and integrated HGSOC scRNA-seq atlas can be downloaded from Zenodo (https://zenodo.org/records/23737155).[85]

Previously published datasets were obtained from the following repositories. The MSK-IMPACT 50K cohort[38] was accessed through Zenodo (https://doi.org/10.5281/zenodo.18445440)[86] and cBioPortal; MSK-MET[43] and the immunotherapy cohort of Samstein et al.[44] through cBioPortal; TCGA and CPTAC data through the Genomic Data Commons and UCSC Xena; and PCAWG[45] consensus data through the ICGC open-access tier. Hartwig-derived WGD calls and gain-timing posteriors were obtained from the GRITIC data release[6] (https://doi.org/10.5281/zenodo.12010145).[87] TRACERx[47] processed data were obtained from Zenodo (https://doi.org/10.5281/zenodo.7822002 and https://doi.org/10.5281/zenodo.7683605)[88,89] and TRACERx–PEACE[8] processed data from Zenodo (https://doi.org/10.5281/zenodo.19737930)[90]; the corresponding raw sequencing data are held in the European Genome–Phenome Archive (EGAS00001008217; controlled access). Processed single-cell copy-number data were obtained from Zenodo (https://doi.org/10.5281/zenodo.13754279 and https://doi.org/10.5281/zenodo.6998936)[91,92];

SPECTRUM[7] data from Synapse (syn66366718) and the Gene Expression Omnibus (GSE180661); and chromatin-compartment and related genomic datasets from Zenodo (https://doi.org/10.5281/zenodo.7351767 and https://doi.org/10.5281/zenodo.7351776)[93,94]. Accession numbers, source links and access conditions for all datasets are provided in **Supplementary Table 13**.

**Methods References**

## Acknowledgements

We thank the patients and their families for their participation and the clinical teams for their support. We thank the Bioinfo Composer team for providing access to the *inseq* high-performance computing platform used in this study. This work was supported by the National Natural Science Foundation of China (82272898, 82203723, 82271940 and 82471932), the Three-Year Action Plan to Promote Clinical Skills and Clinical Innovation Capacity of Municipal Hospitals by the Shanghai Shenkang Hospital Development Center (SHDC2020CR5003-001, SHDC2025CCS006 and SHDC2026CCS006), the Science and Technology Commission of Shanghai Municipality (21ZR1415000), the CSCO-CYH Oncology Research Fund (Y-Young2024-0297), and the Autonomous Region Key Research and Development Program (2024B03037-1).

## Author contributions

Y.Y.Z., X.H.W. and Q.H.G. conceived the study. Y.Y.Z. developed the computational methods, performed the analyses and prepared the figures. X.Y.Z. and Z.L.W. contributed to methodological development, with Z.L.W. also contributing to interpretation of the results. H.M.L. characterized miliary peritoneal dissemination and curated the associated clinical data. Y.Q.C., S.C.M., J.M.L. and S.M.Y. collected, processed and curated clinical and molecular data. Y.Y.Z. wrote the original manuscript. X.W., S.W., X.H.W. and Q.H.G. reviewed and edited the manuscript, supervised the study and acquired funding. All authors read and approved the final manuscript.

## Competing interests

The authors declare no competing interests.

## Extended Data Figures

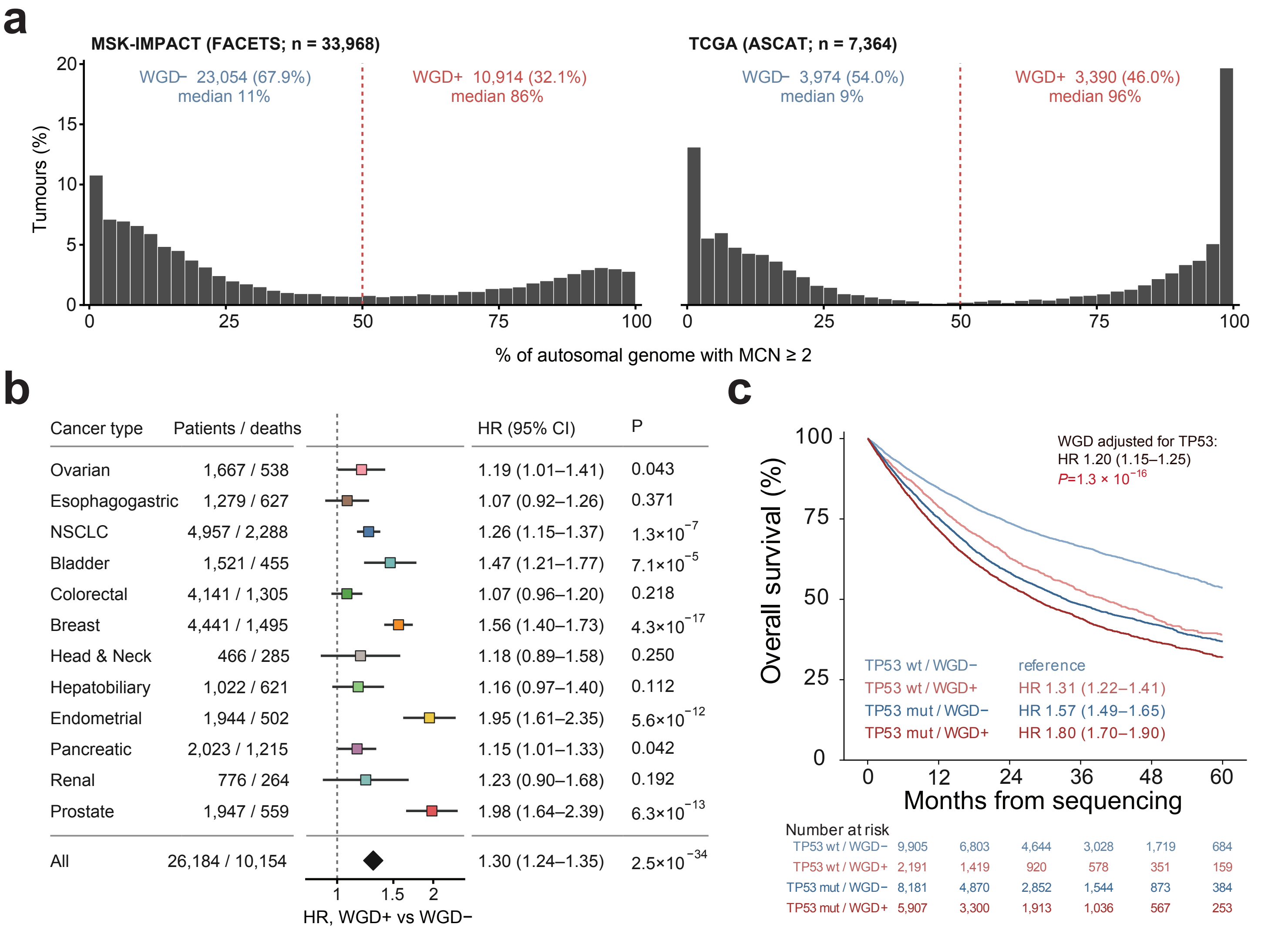
a
MSK-IMPACT (FACETS; n = 33,968)
WGD− 23,054 (67.9%)
median 11%
WGD+ 10,914 (32.1%)
median 86%
TCGA (ASCAT; n = 7,364)
WGD− 3,974 (54.0%)
median 9%
WGD+ 3,390 (46.0%)
median 96%
Tumours (%)
0
5
10
15
20
0
25
50
75
100
% of autosomal genome with MCN ≥ 2
b
Cancer type
Patients / deaths
HR (95% CI)
P
Ovarian 1,667 / 538 1.19 (1.01–1.41) 0.043
Esophagogastric 1,279 / 627 1.07 (0.92–1.26) 0.371
NSCLC 4,957 / 2,288 1.26 (1.15–1.37) $1.3\times10^{-7}$
Bladder 1,521 / 455 1.47 (1.21–1.77) $7.1\times10^{-5}$
Colorectal 4,141 / 1,305 1.07 (0.96–1.20) 0.218
Breast 4,441 / 1,495 1.56 (1.40–1.73) $4.3\times10^{-17}$
Head & Neck 466 / 285 1.18 (0.89–1.58) 0.250
Hepatobiliary 1,022 / 621 1.16 (0.97–1.40) 0.112
Endometrial 1,944 / 502 1.95 (1.61–2.35) $5.6\times10^{-12}$
Pancreatic 2,023 / 1,215 1.15 (1.01–1.33) 0.042
Renal 776 / 264 1.23 (0.90–1.68) 0.192
Prostate 1,947 / 559 1.98 (1.64–2.39) $6.3\times10^{-13}$
All 26,184 / 10,154 1.30 (1.24–1.35) $2.5\times10^{-34}$
1
1.5
2
HR, WGD+ vs WGD−
c
WGD adjusted for TP53:
HR 1.20 (1.15–1.25)
$P$=$1.3\times10^{-16}$
Overall survival (%)
0
25
50
75
100
TP53 wt / WGD− reference
TP53 wt / WGD+ HR 1.31 (1.22–1.41)
TP53 mut / WGD− HR 1.57 (1.49–1.65)
TP53 mut / WGD+ HR 1.80 (1.70–1.90)
0
12
24
36
48
60
Months from sequencing
Number at risk
TP53 wt / WGD− 9,905 6,803 4,644 3,028 1,719 684
TP53 wt / WGD+ 2,191 1,419 920 578 351 159
TP53 mut / WGD− 8,181 4,870 2,852 1,544 873 384
TP53 mut / WGD+ 5,907 3,300 1,913 1,036 567 253

**Extended Data Fig. 1 | WGD classification and survival associations across cancer types.**

**a**, Distribution of the autosomal genome fraction with major copy number ≥2 in MSK-IMPACT (*n* = 33,968) and TCGA (*n* = 7,364); dashed lines indicate the WGD classification threshold. **b**, Cancer-type-specific and pooled hazard ratios (HRs) for overall survival from first sequencing in MSK-IMPACT (*n* = 26,184). Squares indicate cancer-type-specific estimates; diamond, pooled estimate; bars, 95% confidence intervals (CIs). Cox models were stratified by disease state at sequencing, with additional cancer-type stratification for the pooled estimate. **c**, Overall survival by *TP53* mutation and WGD status in the same cohort. Kaplan–Meier curves show numbers at risk; HRs are relative to *TP53*-wild-type, WGD− tumours.

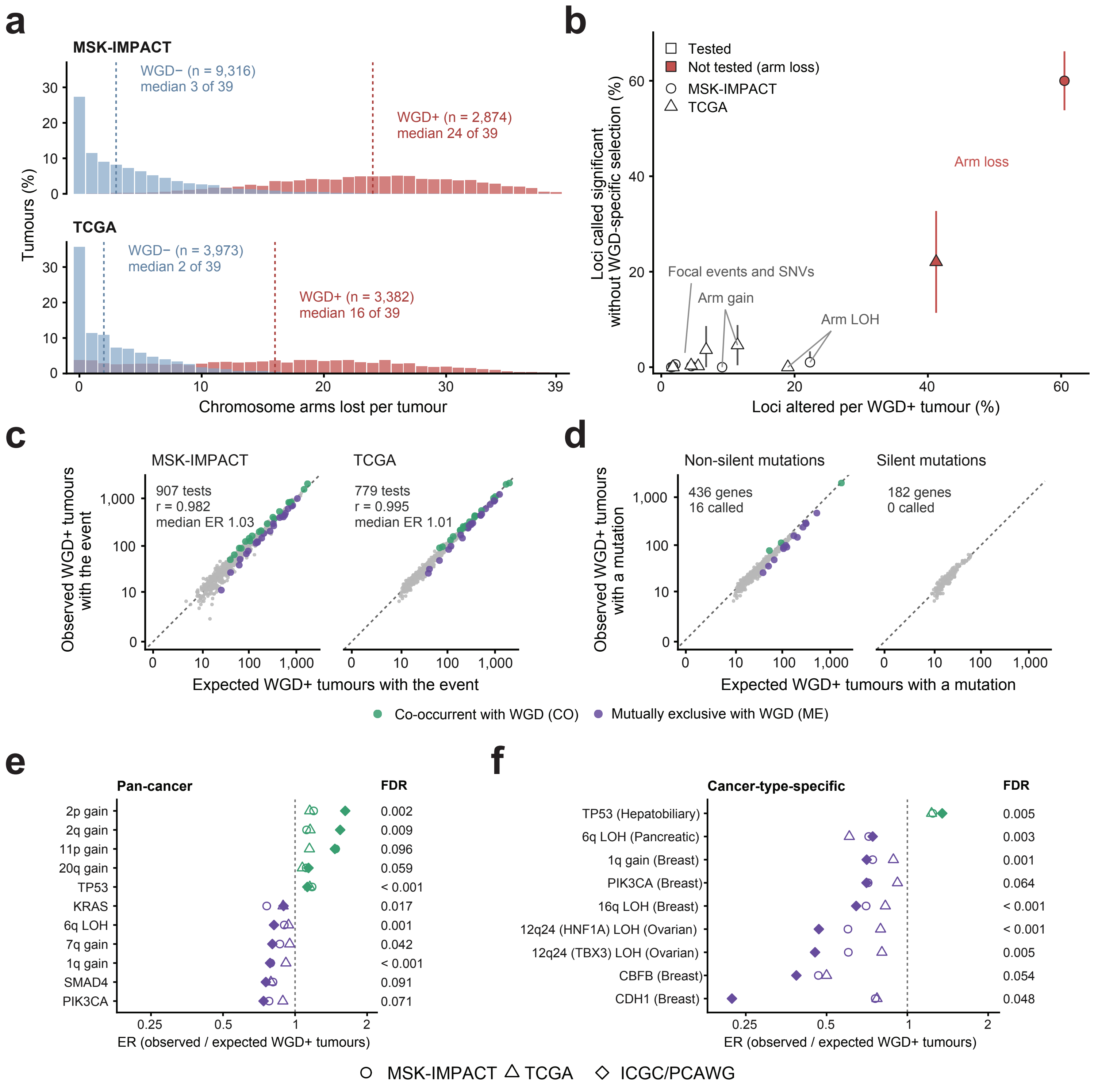

a
MSK-IMPACT
WGD− (n = 9,316)
median 3 of 39
WGD+ (n = 2,874)
median 24 of 39
TCGA
WGD− (n = 3,973)
median 2 of 39
WGD+ (n = 3,382)
median 16 of 39
Tumours (%)
Chromosome arms lost per tumour
b
Tested
Not tested (arm loss)
MSK-IMPACT
TCGA
Arm loss
Focal events and SNVs
Arm gain
Arm LOH
Loci called significant without WGD-specific selection (%)
Loci altered per WGD+ tumour (%)
c
MSK-IMPACT
907 tests
r = 0.982
median ER 1.03
TCGA
779 tests
r = 0.995
median ER 1.01
Observed WGD+ tumours with the event
Expected WGD+ tumours with the event
d
Non-silent mutations
436 genes
16 called
Silent mutations
182 genes
0 called
Observed WGD+ tumours with a mutation
Expected WGD+ tumours with a mutation
Co-occurrent with WGD (CO)
Mutually exclusive with WGD (ME)
e
Pan-cancer
FDR
2p gain 0.002
2q gain 0.009
11p gain 0.096
20q gain 0.059
TP53 < 0.001
KRAS 0.017
6q LOH 0.001
7q gain 0.042
1q gain < 0.001
SMAD4 0.091
PIK3CA 0.071
ER (observed / expected WGD+ tumours)
f
Cancer-type-specific
FDR
TP53 (Hepatobiliary) 0.005
6q LOH (Pancreatic) 0.003
1q gain (Breast) 0.001
PIK3CA (Breast) 0.064
16q LOH (Breast) < 0.001
12q24 (HNF1A) LOH (Ovarian) < 0.001
12q24 (TBX3) LOH (Ovarian) 0.005
CBFB (Breast) 0.054
CDH1 (Breast) 0.048
ER (observed / expected WGD+ tumours)
MSK-IMPACT
TCGA
ICGC/PCAWG

**Extended Data Fig. 2 | Null-model calibration and independent replication of WGD associations.**

**a**, Distribution of chromosome-arm losses per tumour in MSK-IMPACT and TCGA primary tumours; dashed lines indicate medians. **b**, Null-model calibration using five simulated WGD-neutral cohorts preserving tumour-level alteration burden, WGD status and strata. Points indicate simulation means; bars, s.d. Red indicates arm-level losses excluded from association testing. **c**, Observed versus expected numbers of WGD+ tumours carrying each alteration; dashed lines indicate equality. **d**, Negative-control analysis of non-silent and silent mutations in TCGA. **e**,**f**, Independent replication of pan-cancer (**e**) and cancer-type-specific (**f**) WGD associations in ICGC/PCAWG. Symbols distinguish MSK-IMPACT, TCGA and ICGC/PCAWG; green indicates co-occurrence and purple indicates mutual exclusivity with WGD. ER, enrichment ratio; FDR, false discovery rate; LOH, loss of heterozygosity.

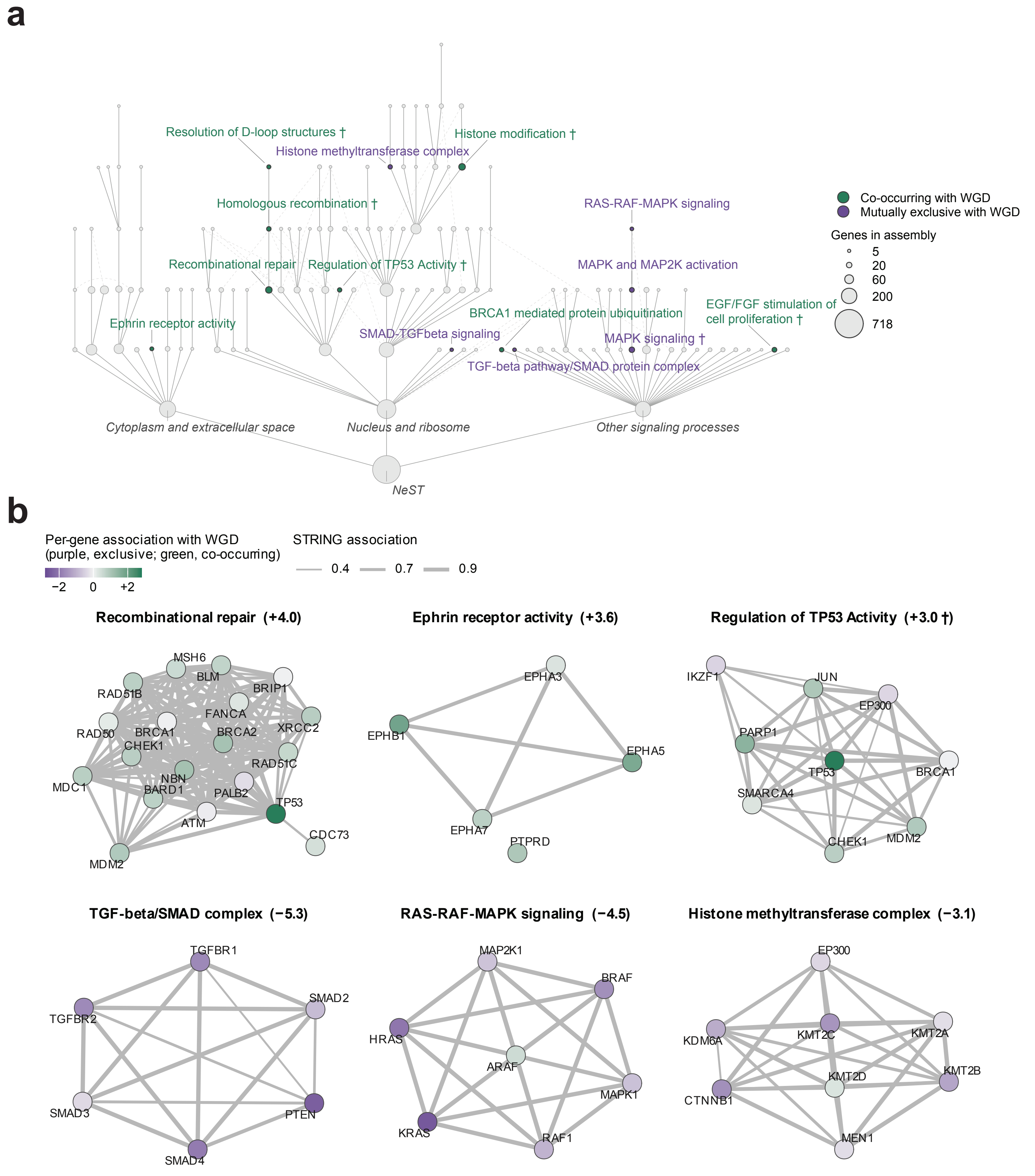
a
Resolution of D-loop structures †
Histone methyltransferase complex
Histone modification †
Homologous recombination †
RAS-RAF-MAPK signaling
Recombinational repair
Regulation of TP53 Activity †
MAPK and MAP2K activation
Ephrin receptor activity
BRCA1 mediated protein ubiquitination
EGF/FGF stimulation of cell proliferation †
SMAD-TGFbeta signaling
MAPK signaling †
TGF-beta pathway/SMAD protein complex
Cytoplasm and extracellular space
Nucleus and ribosome
Other signaling processes
NeST
Co-occurring with WGD
Mutually exclusive with WGD
Genes in assembly
5
20
60
200
718
b
Per-gene association with WGD (purple, exclusive; green, co-occurring)
−2
0
+2
STRING association
0.4
0.7
0.9
Recombinational repair (+4.0)
MSH6
BLM
RAD51B
BRIP1
FANCA
RAD50
BRCA1
BRCA2
XRCC2
CHEK1
RAD51C
NBN
MDC1
BARD1
PALB2
TP53
ATM
CDC73
MDM2
Ephrin receptor activity (+3.6)
EPHA3
EPHB1
EPHA5
EPHA7
PTPRD
Regulation of TP53 Activity (+3.0 †)
IKZF1
JUN
EP300
PARP1
TP53
BRCA1
SMARCA4
CHEK1
MDM2
TGF-beta/SMAD complex (−5.3)
TGFBR1
SMAD2
TGFBR2
SMAD3
PTEN
SMAD4
RAS-RAF-MAPK signaling (−4.5)
MAP2K1
BRAF
HRAS
ARAF
MAPK1
KRAS
RAF1
Histone methyltransferase complex (−3.1)
EP300
KDM6A
KMT2C
KMT2A
KMT2B
KMT2D
CTNNB1
MEN1

**Extended Data Fig. 3 | Protein assemblies associated with WGD.**

**a**, NeST hierarchy showing WGD-associated protein assemblies in MSK-IMPACT and TCGA. Node size indicates assembly size; green, co-occurrence; purple, mutual exclusivity. Filled nodes indicate retained assemblies. Solid edges indicate hierarchical membership; dashed edges indicate additional memberships (pleiotropy). Daggers mark assemblies no longer retained after excluding *TP53*. **b**, Protein-association networks of six representative assemblies. Node fill indicates rank-normalized gene association scores, bounded at ±2.80; edge width indicates STRING protein-association confidence (score ≥ 0.15). Numbers in parentheses indicate assembly-level association scores.

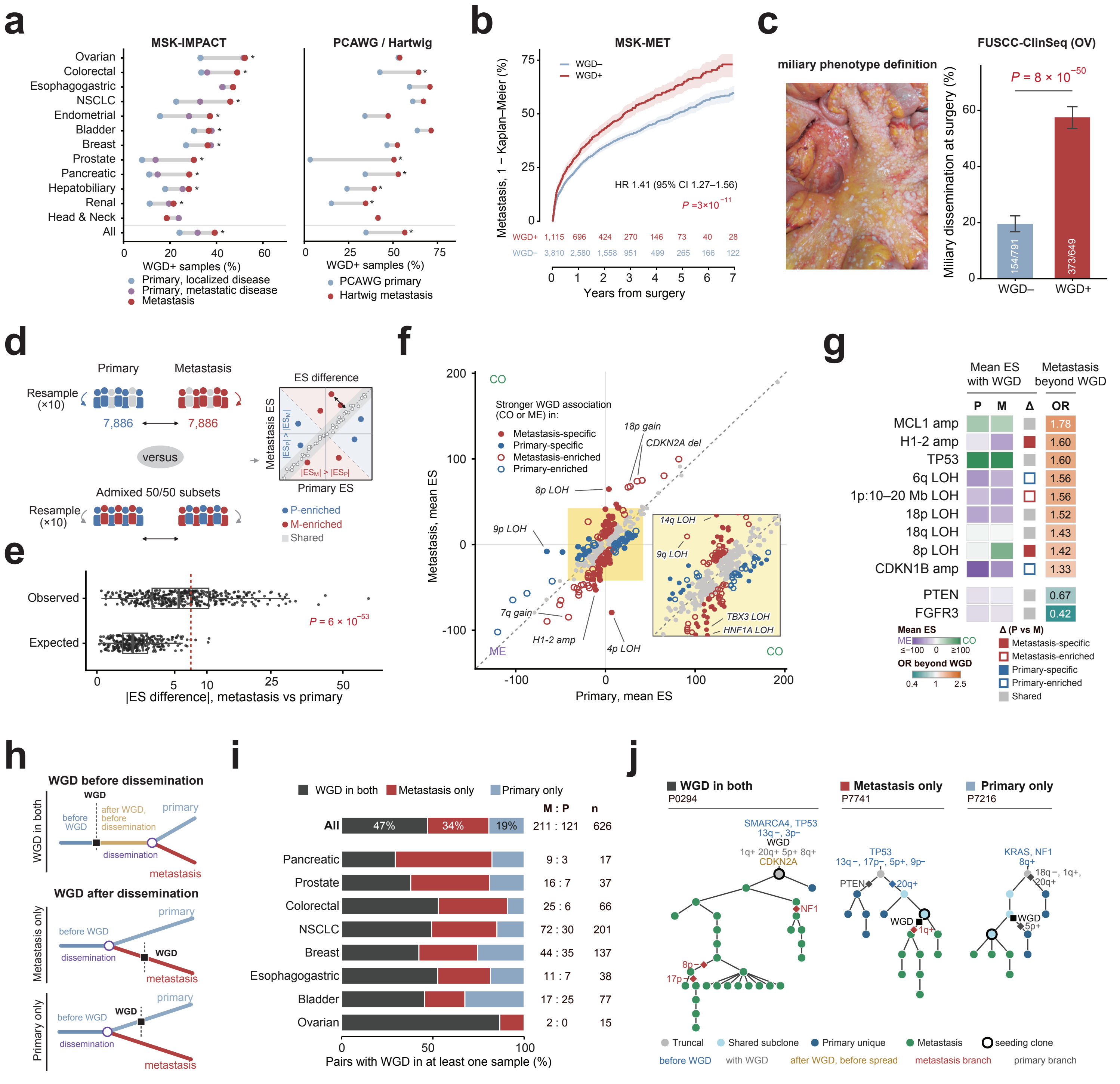

a
MSK-IMPACT
PCAWG / Hartwig
Ovarian
Colorectal
Esophagogastric
NSCLC
Endometrial
Bladder
Breast
Prostate
Pancreatic
Hepatobiliary
Renal
Head & Neck
All
WGD+ samples (%)
Primary, localized disease
Primary, metastatic disease
Metastasis
PCAWG primary
Hartwig metastasis
b
MSK-MET
WGD–
WGD+
Metastasis, 1 – Kaplan–Meier (%)
HR 1.41 (95% CI 1.27–1.56)
P =3×10^−11
WGD+ 1,115 696 424 270 146 73 40 28
WGD– 3,810 2,580 1,558 951 499 265 166 122
Years from surgery
c
FUSCC-ClinSeq (OV)
miliary phenotype definition
P = 8 × 10^−50
Miliary dissemination at surgery (%)
154/791
373/649
WGD–
WGD+
d
Primary
Metastasis
Resample (×10)
7,886
7,886
versus
ES difference
Metastasis ES
Primary ES
|ESP| > |ESM|
|ESM| > |ESP|
Admixed 50/50 subsets
P-enriched
M-enriched
Shared
e
Observed
Expected
P = 6 × 10^−53
|ES difference|, metastasis vs primary
f
CO
ME
Stronger WGD association (CO or ME) in:
Metastasis-specific
Primary-specific
Metastasis-enriched
Primary-enriched
18p gain
CDKN2A del
8p LOH
9p LOH
14q LOH
9q LOH
7q gain
H1-2 amp
4p LOH
TBX3 LOH
HNF1A LOH
Metastasis, mean ES
Primary, mean ES
g
Mean ES with WGD
Metastasis beyond WGD
P
M
Δ
OR
MCL1 amp 1.78
H1-2 amp 1.60
TP53 1.60
6q LOH 1.56
1p:10–20 Mb LOH 1.56
18p LOH 1.52
18q LOH 1.43
8p LOH 1.42
CDKN1B amp 1.33
PTEN 0.67
FGFR3 0.42
Mean ES
ME CO
≤−100 0 ≥100
OR beyond WGD
0.4 1 2.5
Δ (P vs M)
Metastasis-specific
Metastasis-enriched
Primary-specific
Primary-enriched
Shared
h
WGD before dissemination
WGD in both
WGD
before WGD
after WGD, before dissemination
primary
dissemination
metastasis
WGD after dissemination
Metastasis only
before WGD
dissemination
WGD
primary
metastasis
Primary only
before WGD
WGD
dissemination
primary
metastasis
i
WGD in both
Metastasis only
Primary only
M : P
n
All 47% 34% 19% 211 : 121 626
Pancreatic 9 : 3 17
Prostate 16 : 7 37
Colorectal 25 : 6 66
NSCLC 72 : 30 201
Breast 44 : 35 137
Esophagogastric 11 : 7 38
Bladder 17 : 25 77
Ovarian 2 : 0 15
Pairs with WGD in at least one sample (%)
j
WGD in both
P0294
Metastasis only
P7741
Primary only
P7216
SMARCA4, TP53
13q−, 3p−
WGD
1q+ 20q+ 5p+ 8q+
CDKN2A
NF1
8p−
17p−
TP53
13q−, 17p−, 5p+, 9p−
PTEN
20q+
WGD
1q+
KRAS, NF1
8q+
18q−, 1q+, 20q+
WGD
5p+
Truncal
Shared subclone
Primary unique
Metastasis
seeding clone
before WGD
with WGD
after WGD, before spread
metastasis branch
primary branch

**Extended Data Fig. 4 | Independent validation of event timing around WGD.**

**a**, HRD-associated mutagenesis onset relative to WGD in 31 FUSCC-OV tumours; points indicate mutation-time differences; bars, conditional 95% intervals; grey, interval includes zero. **b**, Driver mutation timing in TRACERx lung cancers compared with MSK-IMPACT estimates. **c**, Chromosome-arm loss timing in Hartwig metastases and TRACERx compared with MSK-IMPACT. Blue, pre-WGD; gold, post-WGD; open circles, MSK-IMPACT estimates; solid lines, cohort-specific baselines in **b,c**. **d**, Concordance of arm-loss timing across 39 chromosome arms between MSK-IMPACT and independent cohorts; dashed line, identity. **e**, Probabilities that the first chromosome-arm gain preceded WGD and the last followed WGD in Hartwig metastases (274 arm–cancer-type groups); bars, 95% confidence intervals (CIs); line and band, random-effects mean ± between-group s.d.; dashed line, PCAWG mean. **f**, Single-cell estimates of arm-loss timing from 25 ovarian WGD clone trees compared with ovarian and pan-cancer bulk estimates; point size indicates the number of trees carrying each loss; dashed line, identity. HRD, homologous recombination deficiency.

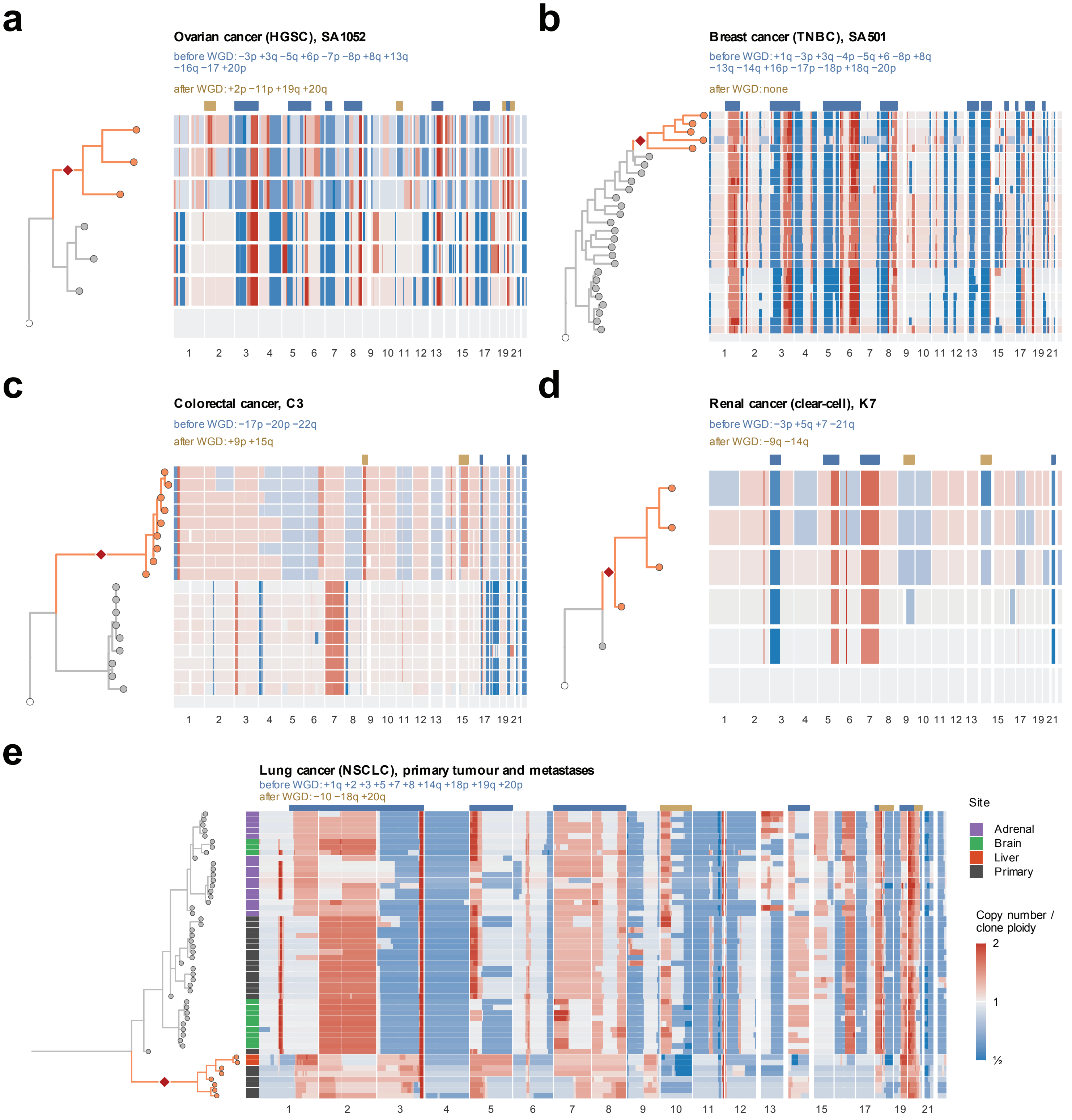

a
Ovarian cancer (HGSC), SA1052
before WGD: −3p +3q −5q +6p −7p −8p +8q +13q −16q −17 +20p
after WGD: +2p −11p +19q +20q
1 2 3 4 5 6 7 8 9 10 11 12 13 15 17 19 21
b
Breast cancer (TNBC), SA501
before WGD: +1q −3p +3q −4p −5q +6 −8p +8q −13q −14q +16p −17p −18p +18q −20p
after WGD: none
1 2 3 4 5 6 7 8 9 10 11 12 13 15 17 19 21
c
Colorectal cancer, C3
before WGD: −17p −20p −22q
after WGD: +9p +15q
1 2 3 4 5 6 7 8 9 10 11 12 13 15 17 19 21
d
Renal cancer (clear-cell), K7
before WGD: −3p +5q +7 −21q
after WGD: −9q −14q
1 2 3 4 5 6 7 8 9 10 11 12 13 15 17 19 21
e
Lung cancer (NSCLC), primary tumour and metastases
before WGD: +1q +2 +3 +5 +7 +8 +14q +18p +19q +20p
after WGD: −10 −18q +20q
1 2 3 4 5 6 7 8 9 10 11 12 13 15 17 19 21
Site
Adrenal
Brain
Liver
Primary
Copy number / clone ploidy
2
1
½

**Extended Data Fig. 5 | Single-cell evidence for the ordering of chromosomal events around WGD.**

**a–e**, Single-cell clone phylogenies and copy-number profiles of high-grade serous ovarian cancer SA1052 (**a**), triple-negative breast cancer SA501 (**b**), colorectal cancer C3 (**c**), clear-cell renal carcinoma K7 (**d**) and non-small-cell lung cancer with primary and metastatic samples (**e**). Grey and orange branches indicate undoubled and doubled clones; diamonds mark WGD events. Blue and gold annotations indicate chromosomal alterations assigned before and after WGD, respectively. Heatmaps show copy number relative to clone ploidy; coloured bars in **e** indicate sampling sites.

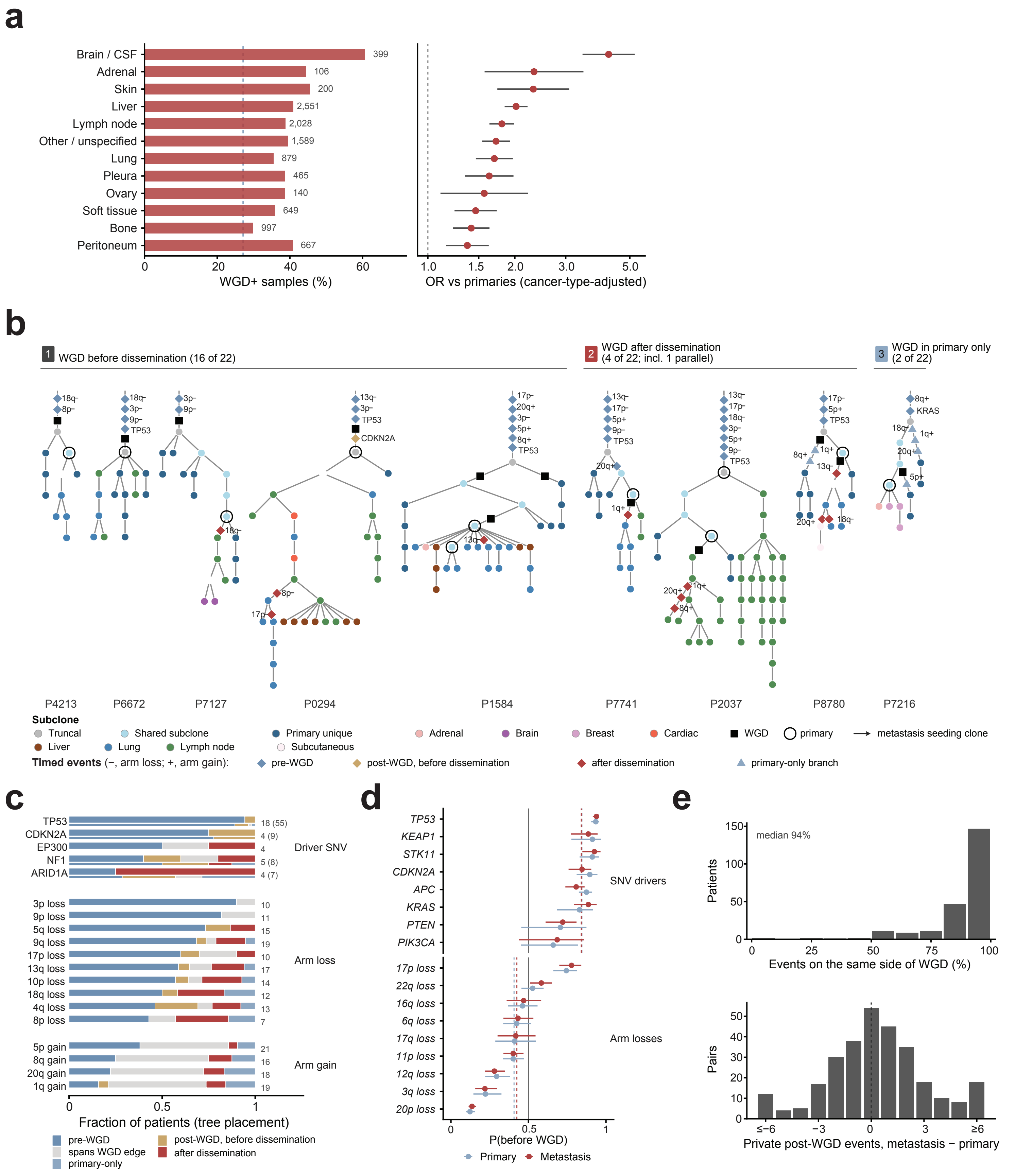
a
Brain / CSF
Adrenal
Skin
Liver
Lymph node
Other / unspecified
Lung
Pleura
Ovary
Soft tissue
Bone
Peritoneum
399
106
200
2,551
2,028
1,589
879
465
140
649
997
667
WGD+ samples (%)
OR vs primaries (cancer-type-adjusted)
b
1 WGD before dissemination (16 of 22)
2 WGD after dissemination (4 of 22; incl. 1 parallel)
3 WGD in primary only (2 of 22)
P4213
P6672
P7127
P0294
P1584
P7741
P2037
P8780
P7216
Subclone
Truncal
Shared subclone
Primary unique
Adrenal
Brain
Breast
Cardiac
WGD
primary
metastasis seeding clone
Liver
Lung
Lymph node
Subcutaneous
Timed events (−, arm loss; +, arm gain):
pre-WGD
post-WGD, before dissemination
after dissemination
primary-only branch
c
TP53
CDKN2A
EP300
NF1
ARID1A
Driver SNV
3p loss
9p loss
5q loss
9q loss
17p loss
13q loss
10p loss
18q loss
4q loss
8p loss
Arm loss
5p gain
8q gain
20q gain
1q gain
Arm gain
Fraction of patients (tree placement)
pre-WGD
post-WGD, before dissemination
spans WGD edge
after dissemination
primary-only
d
TP53
KEAP1
STK11
CDKN2A
APC
KRAS
PTEN
PIK3CA
SNV drivers
17p loss
22q loss
16q loss
6q loss
17q loss
11p loss
12q loss
3q loss
20p loss
Arm losses
P(before WGD)
Primary
Metastasis
e
median 94%
Patients
Events on the same side of WGD (%)
Pairs
Private post-WGD events, metastasis − primary

**Extended Data Fig. 6 | Metastatic distribution of WGD and conservation of evolutionary timing.**

**a**, WGD prevalence across metastatic sites in MSK-IMPACT and cancer-type-adjusted odds ratios (ORs) relative to primary tumours; bars, 95% confidence intervals (CIs); dashed line, primary-tumour prevalence. **b**, Representative TRACERx–PEACE phylogenies showing WGD before or after dissemination, or restricted to primary-tumour lineages. Squares mark WGD; open circles, primary samples; diamonds, genomic events; triangles, primary-only branches. **c**, Timing of driver mutations and chromosomal alterations relative to WGD and dissemination in 22 phylogenies. Colours distinguish evolutionary intervals; thin bars indicate independent driver-timing estimates. **d**, Pre-WGD timing probabilities for driver mutations and chromosome-arm losses in primary and metastatic MSK-IMPACT tumours; bars, 95% CIs. **e**, Distributions of event-timing agreement and differences in private post-WGD events between matched primary–metastasis pairs with WGD at both sites.

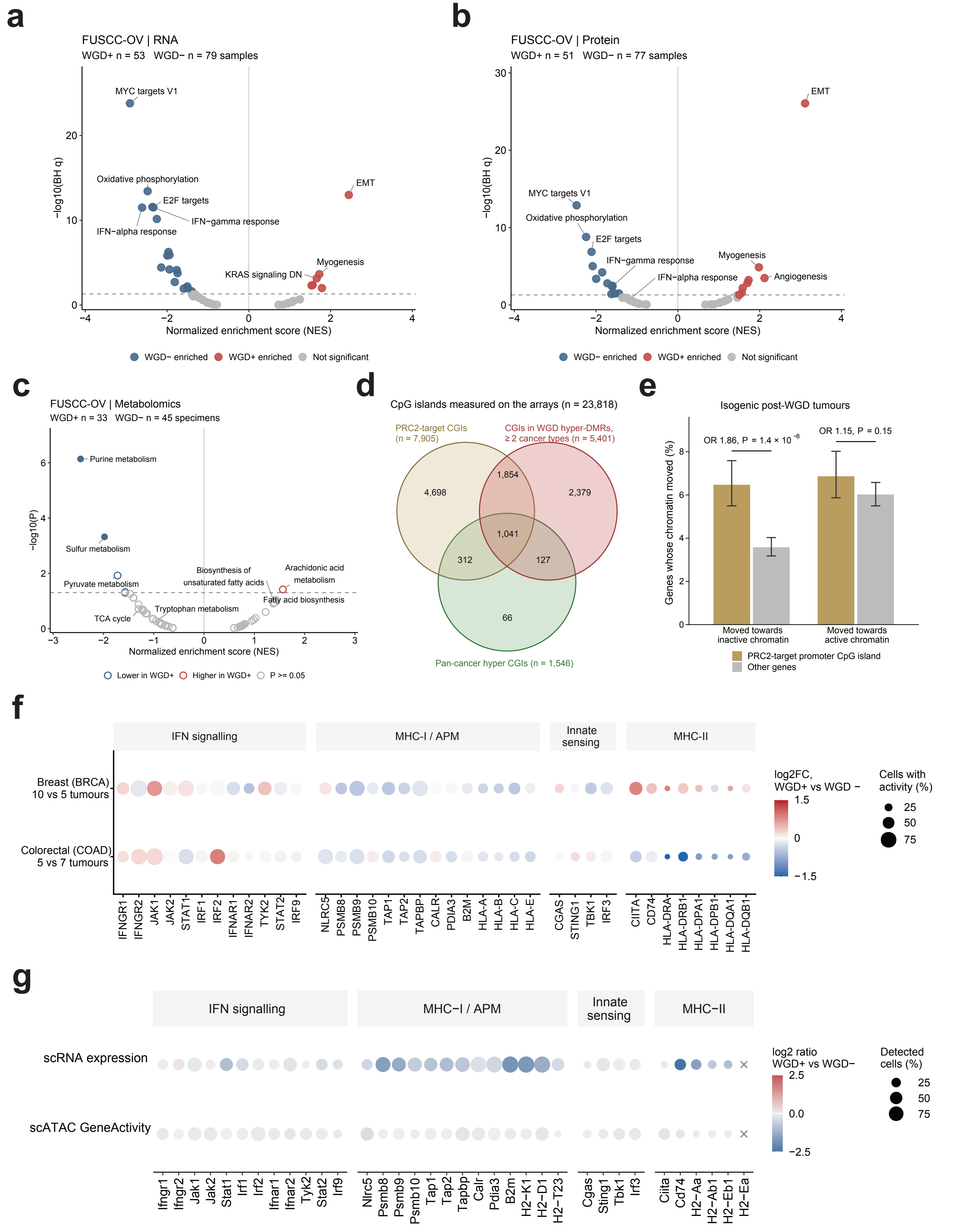

a
FUSCC-OV | RNA
WGD+ n = 53 WGD− n = 79 samples
MYC targets V1
Oxidative phosphorylation
E2F targets
IFN−gamma response
IFN−alpha response
EMT
KRAS signaling DN
Myogenesis
−log10(BH q)
Normalized enrichment score (NES)
WGD− enriched
WGD+ enriched
Not significant
b
FUSCC-OV | Protein
WGD+ n = 51 WGD− n = 77 samples
EMT
MYC targets V1
Oxidative phosphorylation
E2F targets
IFN−gamma response
IFN−alpha response
Myogenesis
Angiogenesis
−log10(BH q)
Normalized enrichment score (NES)
WGD− enriched
WGD+ enriched
Not significant
c
FUSCC-OV | Metabolomics
WGD+ n = 33 WGD− n = 45 specimens
Purine metabolism
Sulfur metabolism
Pyruvate metabolism
Biosynthesis of unsaturated fatty acids
Arachidonic acid metabolism
Fatty acid biosynthesis
Tryptophan metabolism
TCA cycle
−log10(P)
Normalized enrichment score (NES)
Lower in WGD+
Higher in WGD+
P >= 0.05
d
CpG islands measured on the arrays (n = 23,818)
PRC2-target CGIs (n = 7,905)
CGIs in WGD hyper-DMRs, ≥ 2 cancer types (n = 5,401)
4,698
1,854
2,379
1,041
312
127
66
Pan-cancer hyper CGIs (n = 1,546)
e
Isogenic post-WGD tumours
OR 1.86, P = 1.4 × 10^−8
OR 1.15, P = 0.15
Genes whose chromatin moved (%)
Moved towards inactive chromatin
Moved towards active chromatin
PRC2-target promoter CpG island
Other genes
f
IFN signalling
MHC-I / APM
Innate sensing
MHC-II
Breast (BRCA) 10 vs 5 tumours
Colorectal (COAD) 5 vs 7 tumours
log2FC, WGD+ vs WGD −
1.5
0
−1.5
Cells with activity (%)
25
50
75
IFNGR1 IFNGR2 JAK1 JAK2 STAT1 IRF1 IRF2 IFNAR1 IFNAR2 TYK2 STAT2 IRF9
NLRC5 PSMB8 PSMB9 PSMB10 TAP1 TAP2 TAPBP CALR PDIA3 B2M HLA-A HLA-B HLA-C HLA-E
CGAS STING1 TBK1 IRF3
CIITA CD74 HLA-DRA HLA-DRB1 HLA-DPA1 HLA-DPB1 HLA-DQA1 HLA-DQB1
g
IFN signalling
MHC−I / APM
Innate sensing
MHC−II
scRNA expression
scATAC GeneActivity
log2 ratio WGD+ vs WGD−
2.5
0.0
−2.5
Detected cells (%)
25
50
75
Ifngr1 Ifngr2 Jak1 Jak2 Stat1 Irf1 Irf2 Ifnar1 Ifnar2 Tyk2 Stat2 Irf9
Nlrc5 Psmb8 Psmb9 Psmb10 Tap1 Tap2 Tapbp Calr Pdia3 B2m H2−K1 H2−D1 H2−T23
Cgas Sting1 Tbk1 Irf3
Ciita Cd74 H2−Aa H2−Ab1 H2−Eb1 H2−Ea

**Extended Data Fig. 7 | Immune programmes and epigenetic changes associated with WGD.**

**a,b**, Hallmark enrichment in FUSCC-OV transcriptomes (**a**) and proteomes (**b**). Blue and red indicate enrichment in WGD− and WGD+ tumours, respectively; dashed lines, FDR = 0.05. **c**, KEGG pathway enrichment in FUSCC-OV metabolomes; filled points, FDR < 0.05; open points, nominal $P < 0.05$; dashed line, $P = 0.05$. **d**, Overlap of PRC2-target CpG islands, recurrent WGD-associated hypermethylated regions and pan-cancer hypermethylated CpG islands. **e**, Chromatin-compartment changes in isogenic post-WGD tumours, comparing genes with PRC2-target promoter CpG islands and other genes; bars, 95% confidence intervals. **f**, Single-cell ATAC-seq-derived immune-gene activity in breast and colorectal cancers by WGD status. **g**, Single-cell RNA expression and ATAC gene activity in isogenic WGD+ and WGD− mouse tumours at day 14. In **f,g**, colour indicates $\log_2$ fold change or ratio (WGD+ versus WGD−), and point size indicates the fraction of cells with detectable signal; crosses in **g** indicate unscored genes.